\documentclass[prd,twocolumn,showpacs,floatfix,superscriptaddress,nofootinbib]{revtex4}
\usepackage[utf8]{inputenc}
\usepackage[normalem]{ulem}
\usepackage{graphicx}
\usepackage{epsfig}
\usepackage{bm}
\usepackage{amsfonts}
\usepackage[T1]{fontenc}
\usepackage{amssymb}
\usepackage{float}
\usepackage{amsmath}
\usepackage{dcolumn}
\usepackage{wasysym}
\usepackage{cancel}
\usepackage[colorlinks]{hyperref}
\usepackage[usenames,dvipsnames]{color}
\hypersetup{
     breaklinks=true,
    pdfstartview={FitH},    
    colorlinks=true,       
    linkcolor=blue,          
    citecolor=red,        
    filecolor=magenta,      
    urlcolor=blue,           
    anchorcolor=green,      
    linktocpage=true
}

\begin{document}

\title{ Timelike Geodesics of Regular Black Holes 
with Scalar Hair}

\author{P. A. Gonz\'{a}lez}
\email{pablo.gonzalez@udp.cl} \affiliation{Facultad de
Ingenier\'{i}a y Ciencias, Universidad Diego Portales, Avenida Ej\'{e}rcito
Libertador 441, Casilla 298-V, Santiago, Chile.}
\author{Marco Olivares}
\email{marco.olivaresr@mail.udp.cl}
\affiliation{Facultad de
Ingenier\'{i}a y Ciencias, Universidad Diego Portales, Avenida Ej\'{e}rcito
Libertador 441, Casilla 298-V, Santiago, Chile.}
\author{Eleftherios Papantonopoulos}
\email{lpapa@central.ntua.gr}
\affiliation{Physics Division, School of Applied Mathematical and Physical Sciences, National Technical University of Athens, 15780 Zografou Campus, Athens, Greece.}
\author{Yerko V\'{a}squez}
\email{yvasquez@userena.cl}
\affiliation{Departamento de F\'{\i}sica, Facultad de Ciencias, Universidad de La Serena,\\
Avenida Cisternas 1200, La Serena, Chile.}

\begin{abstract}

We investigate timelike geodesics in asymptotically flat regular black holes supported by a phantom scalar field characterized by a scalar charge $A$. This parameter removes the central singularity and continuously deforms the Schwarzschild geometry while preserving asymptotic flatness. We derive the equations of motion for massive test particles and classify bounded and unbounded trajectories in terms of the conserved energy and angular momentum. We determine circular and critical orbits, including the innermost stable circular orbit (ISCO), and analyze the transition between capture and scattering. We show that the scalar charge modifies the location of the unstable and stable circular orbits, the ISCO, and the threshold angular momentum for scattering, exhibiting a nontrivial dependence on the radial coordinate. Their physical scales are naturally described in terms of the invariant areal radius $R(r)=\sqrt{r^2+A^2}$. In the weak-field regime, we compute the perihelion precession and obtain corrections
proportional to $A^2$,
allowing us to constrain the scalar charge from Solar System observations. We also analyze the motion with vanishing angular momentum and show that, while the qualitative structure of the trajectories remains connected to the Schwarzschild limit $A\to 0$, the quantitative deviations encode the geometric effects of the scalar hair.

\end{abstract}

\maketitle

\tableofcontents

\section{Introduction}

Recent cosmological observations indicate that in early cosmological evolution the formation of  matter was governed by  a peculiar matter, the dark energy, which was characterized by negative values of the pressure to density ratio $w$, which could even be $w < -1$ \cite{steinhardt,tegmark,seljak,hannestad,star03,chandra}. A dark energy to produce a negative value of $w$ should be parametrized by a  phantom field having a negative kinetic energy \cite{sen,gorini,fara05}. However, this kind of  perfect-fluid description of dark energy is plagued by instabilities at small scales. To avoid this instability, a phantom scalar field may be regarded as an effective field description following from an underlying theory with positive energies \cite{no03,trod}.

In a recent paper \cite{Farrah:2023opk} it was claimed that regular black holes (BHs) can be supported by astrophysical and cosmological observations as realistic astrophysical BH models, which can become cosmological at a large distance from the BH. Recently, a preferential growth of BH was found from a study of supermassive BHs within elliptical galaxies relative to the stellar mass of the galaxy \cite{akiyama2022first}.  In this way non-singular cosmological BH models can couple to the expansion of the universe, gaining mass proportional to the scale factor. This leads to a realistic behavior at infinity of BH models predicting that the gravitating mass of a BH can increase with the expansion of the universe,  in a manner that depends on the interior solution of the BH. To explain the onset of an accelerating expansion of the universe, it was proposed in \cite{Farrah:2023opk} that stellar remnant BHs are the astrophysical origin of dark energy, explaining the onset of an accelerating expansion of the universe.

Regular BH solutions with an expanding asymptotically de Sitter Kantowski-Sachs  cosmology beyond the event horizon, were found in \cite{Bronnikov:2005gm}. Their solutions were found considering a gravity theory with a scalar field  minimally coupled to gravity with
arbitrary potentials and negative kinetic energy with the purpose of finding regular BH solutions in this spacetime. In the literature there are other solutions, alternative to known ones, with a regular center~\cite{dym92,ned01,bdd03, Barrientos:2025rjn}.

In \cite{Karakasis:2023hni} a gravity theory in the presence of a self-interacting phantom
scalar field minimally coupled to gravity was studied. Local solutions without gravitational singularities, which are points or regions of spacetime where gravitational theory ceases to hold, were found. The singularities are located at the center of the coordinate system, where curvature invariants possess a divergence, and the spacetime is geodesically incomplete. However, Penrose showed that any singularity has to be covered by an event horizon known as the cosmic censorship hypothesis.  In particular, a phantom scalar field with a scalar charge $A$ was considered in gravity theory in \cite{Karakasis:2023hni}. If the scalar charge is zero, then the gravitational singularity is covered by a horizon, as expected. However, if $A$ is not zero, then the scalar charge of the phantom scalar field deforms the geometry in such a way that the gravitational singularity is absent and a compact object is generated with a horizon, which is a regular BH.  It was found that the  charge of the scalar field is connected to the mass of the BH dressing the BH with a secondary hair. The regular black hole solution is based on the phantom scalar configurations originally analyzed in \cite{Bronnikov:2005gm}, where the corresponding energy conditions were also discussed, and  where all the possible solutions were presented. Some properties of this solution like the gravitational lensing and the accretion process can be found in \cite{Ding:2013vta, Ditta2020}. The stability of the background geometry was analyzed in Ref.~\cite{Bronnikov:2012ch} within the linear approximation, where it was shown that these configurations are generically unstable under spherically symmetric perturbations, except for a special class of solutions in which the event horizon coincides with the minimum of the areal radius; the analysis includes both axial perturbations and the monopole sector of polar perturbations. The propagation of a test scalar field on these fixed backgrounds was studied in Ref. \cite{Becar:2026zni}, where the analysis encompasses both the special class of configurations identified in Ref.~\cite{Bronnikov:2012ch} as stable, as well as cases in which the minimum of the areal function lies inside the horizon. Within this probe approximation, it was  shown that there is no evidence of linear instabilities in the scalar sector for any configuration considered, including those that are unstable under gravitational perturbations. This indicates that, at the level of test-field dynamics, the spacetimes support well-defined quasinormal ringing and late-time behavior without the presence of growing modes.

 Important information about the structure and properties of a black hole is given by the study of the motion of a test massive particle around the black hole horizon in particular exploring the generated geodesic
structure. In this way the analysis of the geodesic structure allows us to constrain the parameters appearing in the metric function, and using Solar System observations, we can test the viability of the theory under study \cite{Carvajal:2025ucx,Gonzalez:2020vzl,Gonzalez:2018zuu,Gonzalez:2015jna}.

As discussed above, the particular phantom-scalar regular black-hole
geometry considered in the present work has already been investigated
from several complementary perspectives, including its background
properties, stability, gravitational lensing, accretion, and test-field
dynamics. In a previous work \cite{Gonzalez:2025yjm}, we further
studied this same geometry by analyzing null geodesics and their
weak- and strong-field observational signatures, including Solar
System tests based on light propagation, photon-sphere instability,
and black-hole shadow observables confronted with Event Horizon
Telescope data.

Geodesic motion has also been extensively studied in other black-hole
geometries endowed with scalar hair. Relevant examples can be found
in Refs.~\cite{Gonzalez:2015jna,Ramos:2021jta,
Heydari-Fard:2021pjc,Theodosopoulos:2023ice,Chen:2024luw,
Nekouee:2025zvp}. Related investigations in Horndeski black holes
have been reported in
Refs.~\cite{Carvajal:2025ucx,Carvajal:2025emj,Gonzalez:2020vzl}.
A complementary line of research concerns particle motion in other
regular black-hole geometries, as discussed in
Refs.~\cite{Abbas:2014oua,Stuchlik:2014qja,Azam:2017adt,
Azam:2017izk,Becerril:2020fek,Zhou:2022yio,
Bautista-Olvera:2019blb,Xi:2023oib}. Further related investigations
include the study of geodesic motion, circular orbits, and accretion
around a nonlinear electrodynamics black hole
\cite{Ditta:2023NED}, as well as the analysis of particle dynamics
near a time-conformal slowly rotating Kerr black hole
\cite{Shahzad:2022TC}. Although these backgrounds are physically
different from the phantom-scalar geometry considered here, they
provide useful comparisons for assessing how modifications of the
black-hole spacetime affect orbital structure and particle dynamics.

Within this broader context, the present work provides a systematic
analysis of the timelike geodesic structure of the phantom-scalar
regular black hole. In particular, we extend previous investigations
of this geometry beyond the null-geodesic and circular-orbit sectors
by classifying both bounded and unbounded massive-particle
trajectories. We analyze stable and unstable circular orbits,
planetary motion, second-kind trajectories, and critical
configurations, determine the innermost stable circular orbit (ISCO)
and the threshold separating capture from scattering, and study the
purely radial sector with vanishing angular momentum in terms of
proper and coordinate time. In addition, we derive the correction
induced by the scalar charge $A$ to the perihelion precession and use
Solar System observations to obtain an independent weak-field
constraint on this parameter. Finally, by discussing the
characteristic orbital scales in terms of both the coordinate radius
$r$ and the invariant areal radius
$R(r)=\sqrt{r^2+A^2}$, we distinguish genuine geometric effects of
the scalar hair from coordinate-dependent displacements.

The paper is organized as follows. In Sec.~\ref{setup}, we present the theoretical framework describing regular black holes and summarize the main properties of the corresponding spacetime geometry. Sec.~\ref{geodesics} is devoted to the derivation of the equations of motion for timelike geodesics. In Sec.~\ref{LN0}, we analyze the motion of massive test particles with nonvanishing angular momentum, including bounded and unbounded trajectories, circular and planetary orbits, critical configurations, and the transition between capture and scattering. Sec.~\ref{L0} focuses on the motion with vanishing angular momentum and discusses the corresponding bounded and unbounded trajectories. Finally, Sec.~\ref{FR} contains our conclusions.

\section{Setup of the theory of regular black holes}
\label{setup}

We consider a four-dimensional gravitational theory described by the action
\begin{equation} S = \int d^4x \sqrt{-g}\left[ \frac{R}{2\kappa} - \frac{1}{2}f(\phi)\nabla_{\mu}\phi\nabla^{\mu}\phi - V(\phi)\right]~,\end{equation}
which consists of the Einstein--Hilbert term supplemented by a self-interacting scalar field minimally coupled to gravity. Here $R$ is the Ricci scalar and $\kappa=8\pi G$. Throughout this
work, we adopt geometrized units $G=c=1$. In these units, the mass
parameter $m$, the radial coordinates $r$ and $R$, the scalar charge
$A$, and the angular momentum per unit test-particle mass $L$ have
dimensions of length, whereas the conserved specific energy $E$ is
dimensionless.
The function $f(\phi)$ controls the nature of the scalar field, allowing for canonical ($f(\phi)>0$) and phantom ($f(\phi)<0$) configurations.

Varying the action with respect to the metric and the scalar field leads to the field equations
\begin{eqnarray}
&&G_{\mu\nu} =\kappa T_{\mu\nu} ~,\\
&&f(\phi)\Box\phi +\frac{f'(\phi)}{2}\nabla_{\mu}\phi\nabla^{\mu}\phi  = \frac{dV}{d\phi}~,
\end{eqnarray}
where the energy-momentum tensor is given by
\begin{equation}
T_{\mu\nu} = f(\phi) \nabla_{\mu}\phi\nabla_{\nu}\phi - \frac{f(\phi)}{2}g_{\mu\nu}\nabla_{\alpha}\phi\nabla^{\alpha}\phi - V(\phi)~.
\end{equation}
We focus on static and spherically symmetric configurations described by the line element
\begin{equation} ds^2 = - b(r)dt^2 + \cfrac{1}{b(r)}dr^2 + R(r)^2 d\Omega^2 \label{ds}~,\end{equation}
where
\begin{equation} d\Omega^{2} = d\theta^2 + \sin^{2}\theta d\varphi^2, \end{equation}
and
\begin{equation}
R(r) = \sqrt{r^2+A^2}~, \label{defor}
\end{equation}
where $A$ introduces a new length scale associated with the scalar
sector \cite{Bronnikov:2005gm}. Accordingly, the ratio $A/m$ is dimensionless. The radial coordinate $r$ is not an
areal coordinate. Instead, $R(r)$ is the invariant areal radius, so that the area of the symmetry two-spheres is
$\mathcal{A}=4\pi R^2(r)$. Therefore, when discussing the location
of horizons and characteristic particle orbits, it is useful to
distinguish their coordinate position $r$ from their invariant
geometric scale $R$. We
report both quantities for the principal characteristic radii.

In what follows, we restrict our attention to asymptotically flat solutions supported by a phantom scalar field, $f(\phi)=-1$. In this case, the system admits the exact solution
\begin{eqnarray}\nonumber
b(r) &\equiv& \frac{3m}{A^2}r-\frac{r^2}{A^2}+\frac{(4A-6\pi m)(r^2+A^2)}{4\,A^{3}}+\\ \label{bflat0}
&+&\frac{3 m(r^2+A^2)}{\,A^{3}}\arctan{r\over A}\,,
\end{eqnarray}
accompanied by the scalar profile
\begin{eqnarray}\label{d4solution}
\phi(r) &=& \frac{1}{2 \sqrt{\pi }}\tan ^{-1}\left(\frac{r}{A}\right)~,
\end{eqnarray}
and the self-interaction potential
\begin{eqnarray} \label{d4pot}
\nonumber V(\phi) &=& \frac{3m}{16 \pi A^3} \Bigg[-8 \sqrt{\pi } \phi -3 \sin \left(4 \sqrt{\pi } \phi \right)\\
&&+2 \pi - \left(\pi -4 \sqrt{\pi } \phi \right) \cos \left(4 \sqrt{\pi } \phi \right)  \Bigg] ~.
\end{eqnarray}

The location of the event horizon is determined by the condition $b(r)=0$, which must be solved numerically. Insight into its existence can be obtained from the near-origin behavior,
\begin{equation} b(r\to0)\sim \frac{4 A-6 \pi  m}{4 A}+\frac{6 m r}{A^2}-\frac{3 (\pi  m) r^2}{2 A^3}+\mathcal{O}\left(r^3\right)~,\end{equation}
where the leading constant term governs the sign of the lapse function. Requiring the presence of a root in the physical domain $r\geq 0$ imposes the bound
\begin{equation}\label{mAbound}
m \geq  \frac{2 A}{3 \pi }~,
\end{equation}
which establishes a direct relation between the scalar scale $A$ and the minimum mass required for the existence of a horizon.

In the asymptotic regime $r \gg A$, the lapse function admits the expansion
\begin{equation}
b(r) \approx  1-\frac{2m}{r}+\frac{2mA^2}{5\,r^{3}}\equiv \tilde{b}(r)\,,
\label{bflat1} 
\end{equation}
showing that the spacetime approaches the Schwarzschild solution with subleading corrections controlled by $A$. The corresponding approximate horizon can be obtained from this expansion, yielding
\begin{equation}
r_+ =\frac{2m}{3}+\frac{4m}{3}\cos\left( {1\over 3}\arccos\left(1- {27A^2\over 40m^2}\right) \right) \,.
\label{horizon} 
\end{equation}

The parameter $m$ retains its interpretation as the ADM mass, which can be computed using the Abbott-Deser method for asymptotically flat spacetimes
\begin{equation}\label{d4mass}
M = \frac{1}{2}\lim_{r\to\infty} r\left(\frac{1}{b(r)} -1\right) = m~.
\end{equation}
Importantly, this quantity is not modified by the presence of the scalar field, indicating that $A$ encodes an independent degree of freedom.

The metric function $b(r)$ at infinity reads
\begin{eqnarray} 
 b(r\to\infty) &\sim&   1-\frac{2m}{r}+\frac{2mA^2}{5\,r^{3}}
+\mathcal{O}\left(\left(\frac{1}{r}\right)^5\right)\,,
\end{eqnarray}
and the asymptotic behavior of the potential reads
\begin{equation} V(r\to\infty) \sim \frac{A^2 m}{10 \pi  r^5}-\frac{13 \left(A^4 m\right)}{70 \pi  r^7}+\frac{9 A^6 m}{35 \pi  r^9} + \mathcal{O}\left(\left(\frac{1}{r}\right)^{11}\right)~,\end{equation}
which resembles the Schwarzschild black hole with corrections in the structure of spacetime that depend on the parameter $A$. The series expansion for the scalar field at infinity leads to
\begin{align}
\phi(r \to \infty) = \frac{\sqrt{\pi}}{4} - \frac{1}{2\sqrt{\pi}}\, \frac{A}{r} + \mathcal O\left(\frac{A^3}{r^3}\right)\, ,
\end{align}
from which we conclude that $A$ (or, to be precise, $-\frac{A}{2\sqrt{\pi}}$) plays the role of a scalar charge.

\begin{figure}[!h]
	\begin{center}
\includegraphics[width=80mm]{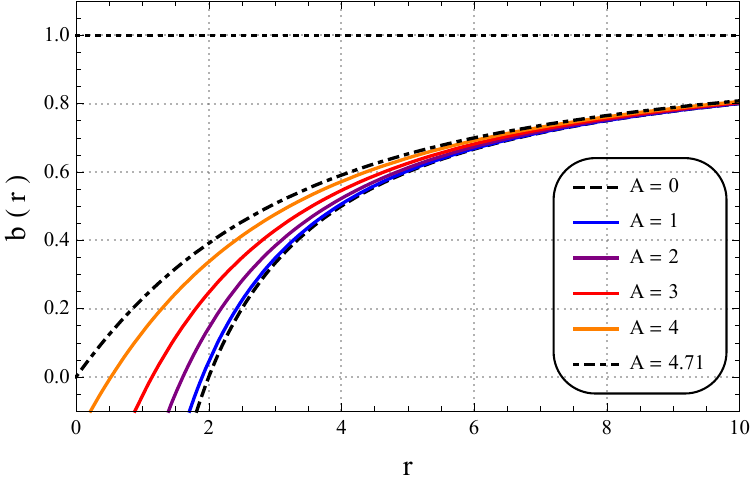}
	\end{center}
	\caption{Lapse function $b(r)$ with $m=1$ and different values of the scalar charge $A$.
		The event horizon is $r_+=2.000$ ($R(r_+)=R_+=2.000$) for $A=0$, $r_+=1.901$ ($R_+=2.148$) for $A=1$, $r_+=1.610$ ($R_+=2.568$) for $A=2$, $r_+=1.146$ ($R_+=3.211$) for $A=3$, $r_+=0.526$ ($R_+=4.034$) for $A=4$, and $r_+=0$ ($R_+=4.710$) for $A=4.710$.}
	\label{f1}
\end{figure}

Fig.~\ref{f1} illustrates the behavior of the lapse function for different values of $A$. In the limit $A\to 0$, the standard Schwarzschild solution is recovered. For nonvanishing $A$, the geometry corresponds to a regular black hole, where the central region is smoothed out. Increasing $A$ shifts the coordinate position of the horizon toward smaller values of $r$; however, the invariant areal radius $R(r_+)$ grows with $A$, indicating that the physical size of the horizon increases. At the critical value $A_{\rm crit}=3\pi m/2$ 
($A_{\rm crit}\simeq4.71$ for $m=1$), the coordinate location of
the horizon reaches $r_+=0$. At this critical configuration, the horizon coincides with the minimum of the areal radius, i.e., with the throat at $r=0$. The critical value of the scalar charge therefore marks a change in the relative location of the throat and the event horizon.
From Eq.~(\ref{mAbound}), the black-hole solutions satisfy
\begin{equation}
\frac{A}{m}\leq \frac{3\pi}{2},
\end{equation}
so that, for the normalization $m=1$ adopted in our numerical
analysis,
\begin{equation}
A_{\rm crit}=\frac{3\pi}{2}\simeq 4.71.
\end{equation}
As $A$ approaches $A_{\rm crit}$ from below, the coordinate
location of the event horizon moves toward $r_+=0$. However, since
the invariant areal radius is $R(r)=\sqrt{r^2+A^2}$, the physical
radius of the limiting surface remains finite,
$R_+=A_{\rm crit}$. Thus, the apparent collapse of the horizon to
$r=0$ is a coordinate statement and does not correspond to a
vanishing invariant size. In this limiting configuration, the
horizon coincides with the minimum of the areal radius, which is
located at $r=0$.

For $A>A_{\rm crit}$, the minimum of the areal radius at $r=0$ lies in the static $R$ region. The zero of the lapse function is then shifted to the $r<0$ side of the extended geometry, i.e., to the other side of the throat. Thus, the horizon moves through the throat from $r>0$ to $r<0$ as $A$ crosses the critical value. Therefore, the asymptotically flat region at $r\to+\infty$ is connected to the throat and the geometry acquires a wormhole-like structure. Consequently, the physical interpretation of geodesics changes
qualitatively across the critical value, since the throat and horizon represent distinct geometric structures. For this reason, the trajectory classification
presented below is restricted to the parameter range
$A\leq A_{\rm crit}$ and its limiting critical configuration. 
The consequences of approaching this critical configuration for the
timelike geodesic structure will be discussed in Sec.~\ref{LN0}, where the
evolution of the circular-orbit scales and the transition between
scattering and capture are analyzed.

In summary, the parameter $A$ encodes a scalar hair associated with the phantom field and plays a dual role: it regularizes the spacetime and introduces nontrivial deviations from Schwarzschild geometry. In the following, we analyze how these modifications affect the dynamics of test particles, focusing on timelike geodesics and their observational implications.

\section{Timelike Geodesics}
\label{geodesics}

In order to obtain a description of the allowed motion of matter in the exterior spacetime of the black hole, we use the standard Lagrangian formalism \cite{Chandrasekhar:579245,Cruz:2004ts,Villanueva:2018kem}, so that the corresponding Lagrangian associated with the line element (\ref{ds}) reads

\begin{equation}\label{Lag1}
\mathcal{L}=-\frac{b(r)\, \dot{t}^2}{2}+ \frac{\dot{r}^2}{2\,b(r)}+\frac{R(r)^2}{2} \left(\dot{\theta}^2+ \sin^2 \theta\,\dot{\varphi} \right) \,,  \end{equation}
where $b(r)$ is the background metric (\ref{bflat0}). Here, the dot indicates differentiation with respect to the proper time $\tau$.  Since the Lagrangian (\ref{Lag1}) does not depend on the coordinates ($t,\varphi$), they are {\it cyclic coordinates} and, therefore, the corresponding conjugate momenta $\pi_{q} = \partial \mathcal{L}/\partial \dot{q}$ are conserved along the geodesic. Explicitly, we have 
\begin{eqnarray}\label{pp1}
\pi_{t}&=& -b(r)\, \dot{t}\equiv -E\,, \\\label{angcons} 
\pi_{\varphi}&=&R(r)^2\sin^2 \theta\, \dot{\varphi}\equiv L\,,
\end{eqnarray}
where $E$ denotes a positive constant associated with the time invariance of the Lagrangian, which can be identified with the conserved energy in the asymptotically flat limit of the spacetime defined by the line element (\ref{ds}). Similarly, 
$L$ corresponds to the conserved angular momentum, which implies that the motion is confined to an invariant plane. In this work, we restrict our analysis to the equatorial plane
$\theta= \pi/2$, such that $\dot\theta =0$.  Accordingly,  from Eq. (\ref{angcons}) we obtain: 
\begin{equation}
 \dot{\varphi}=\frac{L}{R(r)^2}\,.
\label{eq1} 
\end{equation}
Therefore, using the fact that $2\mathcal{L}=-h^2$, where $h$ is the test mass,  together with Eqs. (\ref{pp1}) and (\ref{eq1}), we obtain the following equations of motion 

\begin{eqnarray}
\label{w.12}
&&\left(\frac{{\rm d}r}{{\rm d} \tau}\right)^{2}= E^2-V^2(r)\,,\\
\label{w.13}
&&\left(\frac{{\rm d} r}{{\rm d} t}\right)^{2}= {b^{\,2}(r)\over E^2}\left[E^2-V^2(r)\right]\,,\\
\label{w.14}
&&\left(\frac{{\rm d} r}{{\rm d} \varphi}\right)^{2}=  {R(r)^{\,4}\over L^2}\left[E^2-V^2(r)\right]\,,
\end{eqnarray}
where the effective potential $V^2(r)$ is defined by
\begin{equation}
V^2(r) \equiv b(r)\left[h^2+\frac{L^{2}}{R(r)^{2}}\right]\,.
\label{Veff1} 
\end{equation}
Finally, by normalization, we consider $h=1$ for massive particles. The case $h=0$ (photons) was studied in Ref. \cite{Gonzalez:2025yjm}. Unless otherwise stated, throughout the numerical analysis in
Secs.~\ref{LN0} and \ref{L0} we set $m=1$. Consequently, all quantities with
dimensions of length, including $r$, $R$, $A$, and $L$, are expressed
in units of the black-hole mass parameter $m$.

\section{Motion with $L\neq 0$}
\label{LN0}

Throughout this section, we restrict the analysis to
$A\leq A_{\rm crit}=3\pi m/2$, including the limiting critical
configuration. In the following, the trajectories can be classified into two regimes:
bounded ($E<1$) and unbounded ($E\geq 1$) trajectories.

\subsection{Bounded Orbits ($E<1$)}

\subsubsection{Circular orbits.}

\begin{figure}[!h]
	\begin{center}
\includegraphics[width=65mm]{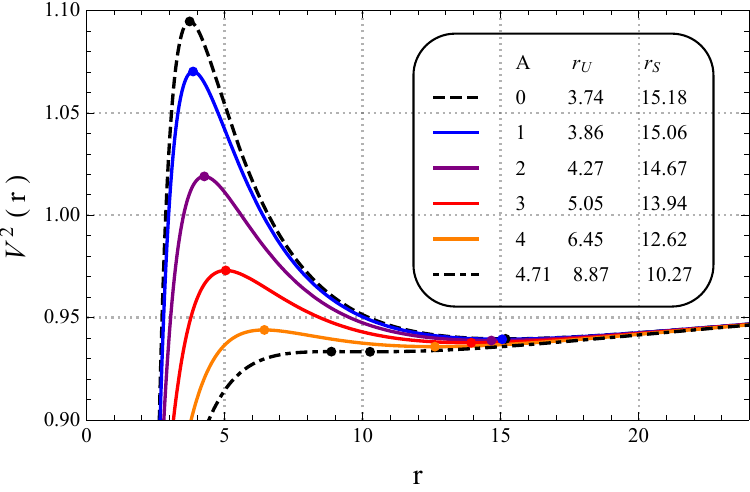}
	\end{center}
	\caption{Effective potential $V^2(r)$ for different values of the scalar charge $A$. Here, $m=1$ and $L=4.35$.}
	\label{potential1}
\end{figure}

The existence of extrema in the effective potential determines the
presence of circular geodesics. In particular, a local maximum
corresponds to an unstable circular orbit ($r_U$), while a local minimum corresponds to a stable one ($r_S$). These radii are obtained from the standard condition $V^2\,'(r)=0$. Using the explicit form of the effective potential given in Eq.~(\ref{Veff1}), this condition leads to
\begin{equation} \label{polyrc}
\frac{3m}{A^3}\left( 2A-\pi \,r+2\,r\arctan{r\over A}\right) -\frac{2(r-3m)L^2}{(r^2+A^2)^2}=0\,.
\end{equation}
Eq.~(\ref{polyrc}) does not admit a closed analytic solution for $r$, and
therefore the radii of the circular orbits must be determined
numerically. The resulting solutions define the unstable radius $r_U$
and the stable radius $r_S$. Fig.~\ref{potential1} displays the effective potential $V^2(r)$ for $m=1$ and
$L=4.35$ for several values of the scalar charge $A$. 
The inset table summarizes the numerical
values of $r_U$ and $r_S$ obtained from Eq.~(\ref{polyrc}). 
As the scalar charge $A$ increases, both the coordinate radius
$r_U$ and the corresponding invariant areal radius $R_U$ of the
unstable circular orbit increase. In contrast, both $r_S$ and
$R_S$ associated with the stable circular orbit decrease.
Consequently, the separation between the stable and unstable
circular orbits becomes progressively smaller. The corresponding
values of both radial quantities are summarized in
Table~\ref{tab:characteristic_radii}. This shows that the opposite
displacements of the stable and unstable circular orbits are not coordinate-dependent features, since the same qualitative behavior
is obtained in terms of the invariant areal radius.
At the same time, the height of the
potential barrier decreases, reflecting a weakening of the effective
gravitational confinement. In the limit $A\rightarrow0$, the standard
Schwarzschild configuration is smoothly recovered.

Once the circular orbit radius $r_c$ is determined,
the orbital motion can be characterized through its proper and coordinate period.
Using Eq.~(\ref{eq1}), the proper period of a particle moving along a circular
orbit of radius $r_c$ is given by
\begin{equation}\label{p1}
T_{\tau}=\frac{2\pi\,(r_{c}^2+A^2)}{L}\,.
\end{equation}
Solving Eq.~(\ref{polyrc}) for $L$ and substituting the result into Eq.~(\ref{p1}), we obtain
\begin{equation}\label{p1b}
T_{\tau}=2\pi\, \sqrt{\frac{ 2A^3(r_{c}-3m)}{3m\left( 2A-\pi \,r_{c}+2\,r_{c}\arctan{r_{c}\over A}\right)}}\,.
\end{equation}
The coordinate period follows from Eq.~(\ref{pp1}) as
\begin{equation}\label{p2}
T_t=2\pi\, \sqrt{\frac{ 2\,r_{c}A^3}{3m\left( 2A-\pi \,r_{c}+2\,r_{c}\arctan{r_{c}\over A}\right) }}\,.
\end{equation}

In the regime $A \ll r$, the metric function $b(r)$ can be approximated
by the expression given in Eq.~(\ref{bflat1}). Substituting this form into the
effective potential Eq.~(\ref{Veff1}), and imposing the circular-orbit condition
$V^{2\,\prime}(r)=0$, we obtain
\begin{eqnarray} \label{polyrc0}
m(5r^2-3A^2)-\frac{(5r^5-15mr^4+3mA^4)L^2}{(r^2+A^2)^2}=0\,.
\end{eqnarray}
Now, solving Eq.~(\ref{eq1}) for the angular momentum and substituting the result
into the expression for the proper period, we obtain
\begin{equation}\label{p10}
T_{\tau}=2\pi\, \sqrt{\frac{5r^5-15mr^4+3mA^4}{m(5r^2-3A^2)}}\,,
\end{equation}
while the corresponding coordinate period reads
\begin{equation}\label{p20}
T_t=2\pi\, \sqrt{\frac{5 r_{c}^5}{m(5r^2-3A^2)}}\,.
\end{equation}	
Setting $A=0$ ($R(r)=r$), these expressions reduce to the well-known Schwarzschild
results
\begin{equation}\label{p10S}
T_{\tau}=2\pi\, \sqrt{\frac{r^3-3mr^2}{m}}\,,
\end{equation}
and 
\begin{equation}\label{p20}
T_t=2\pi\, \sqrt{\frac{r_{c}^3}{m}}\,.
\end{equation}

To determine the epicyclic frequency associated with small radial
perturbations around the stable circular orbit, we expand the effective
potential $V^2(r)$ in a Taylor series about the stable radius
$r=r_S$. This yields
\begin{eqnarray}\label{e17}
V^2(r)=V^2 (r_S)+ V^2 \,' (r_S)(r-r_S)\\+{1\over2}V^2\, ''(r_S)(r-r_S)^2+ \cdots\,, \nonumber
\end{eqnarray}
where $(\cdot)'$ denotes differentiation with respect to the radial
coordinate. Since $r_S$ corresponds to a stable circular orbit,
the first derivative vanishes $V^{2\,\prime}(r_S)=0$. So, by defining the smaller
coordinate $x=r-r_S$, together with the epicycle frequency
$\kappa^2 := V^2 \, ''(r_S)/2 $ \cite{RamosCaro:2011wx},  we can rewrite the above equation as
\begin{equation}\label{e18}
V^2(x)\approx E_S^2+\kappa^2\,x^2\,,
\end{equation}
where ${E_S}:=V(r_S)$ is the energy of the particle in a stable circular orbit.
Moreover, the test particles satisfy the harmonic equation of motion $\ddot{x}=-\kappa^2\,x$, and the epicycle frequency is given by

\begin{eqnarray}\label{e20}\nonumber
\kappa^{2}&=& \frac{3m\left( 3\pi mA^2-2A^3 -r_S(4r_S-15m)(\pi r_S-2A)\right) }
{2 A^3 (r_S-3 m)( r_S^2+A^2)}\\
                &+&\frac{3m(4\,r_S^{3}-15mr_S^{2}-3mA^2)}
                { A^3 (r_S-3 m)( r_S^2+A^2)} \, \arctan{r_S\over A}\,.
\end{eqnarray}

As we mentioned, in the regime $A\ll r$, the metric function $b(r)$ can be approximated by Eq.~(\ref{bflat1}).
The epicycle frequency is given by
\begin{equation}\label{k2Aprox}
\kappa^{2}=\frac{m\left( 25 r^6(r_S-6m)-30r^4A^2 (2r_S-5m)\right) }
{25\, r_S^7 \,( r_S^2+A^2) (r_S-3 m)}\,.
\end{equation}
Finally, setting $A=0$, Eq.~(\ref{k2Aprox}) reduces to the Schwarzschild result

\begin{equation}\label{k2schw}
\kappa^{2}=\frac{m(r-6m)}{r^3(r-3m)}\,.
\end{equation}

\subsubsection{Planetary orbits}

\begin{figure}[!h]
	\begin{center}
\includegraphics[width=65mm]{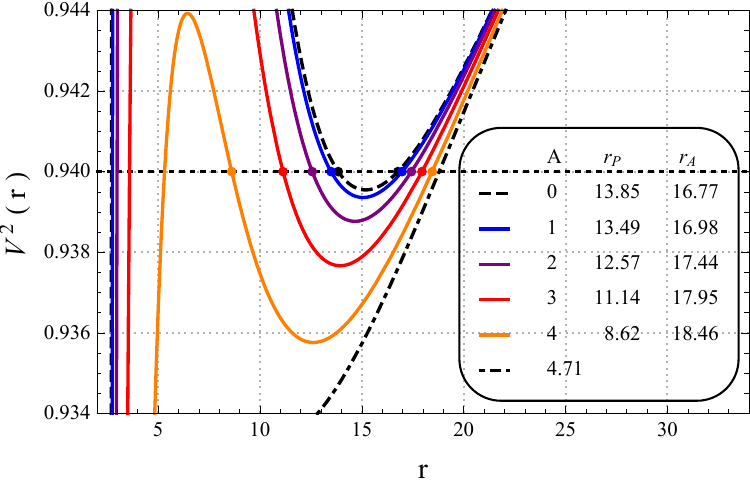}
	\end{center}
	\caption{Effective potential $V^2(r)$ for different values of the scalar charge $A$. Here $r_P$ and $r_A$ are the perihelion and aphelion distances, respectively. Here, $m=1$, $L=4.35$
    and $E^2=0.94$.}
	\label{potential2}
\end{figure}

Planetary orbits can exist for values of the angular momentum 
$L > L_{\text{ISCO}}$, where $L_{\text{ISCO}}$ denotes the angular momentum 
corresponding to the innermost stable circular orbit (ISCO).
Fig.~\ref{potential2} displays the effective potential $V^2(r)$ for fixed parameters $m=1$, $L=4.35$, and energy $E^2=0.94$, for several values of the scalar charge $A$. 
The horizontal line represents $E^2$, and its intersections with the potential determine the radial turning points corresponding to the perihelion $r_P$ and aphelion $r_A$ distances. 
The table in the figure reports the numerical values of $r_P$ and $r_A$ for each $A$. 
As the scalar charge increases, the perihelion distance $r_P$--and $R_P$-- shifts inward while the aphelion distance $r_A$--and $R_A$-- moves outward, leading to a widening of the allowed radial interval. 
This behavior reflects the modification of the effective potential well induced by the scalar hair. 
In the limit $A \to 0$, the Schwarzschild configuration is recovered.

\begin{figure}[H]
	\begin{center}
		\includegraphics[width=4cm]{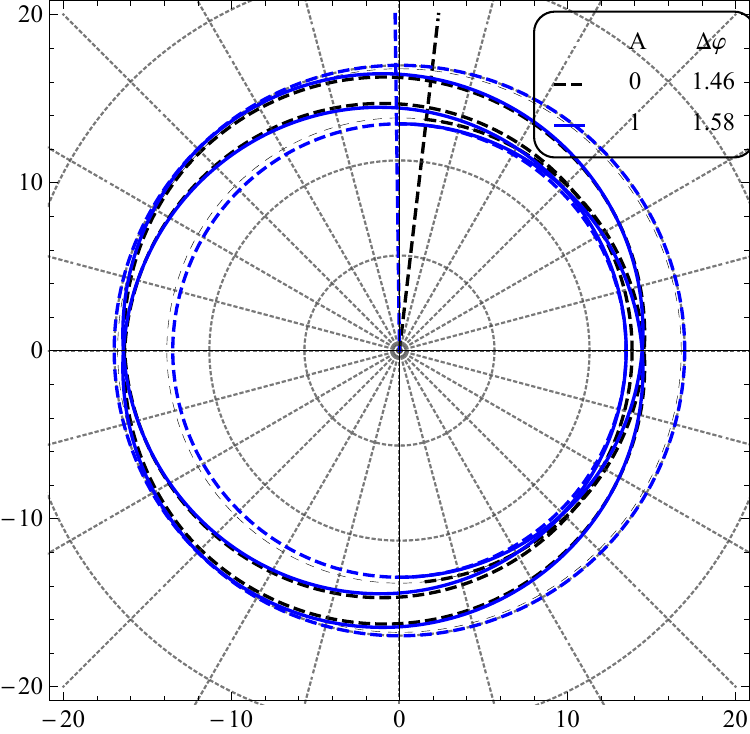}
        \includegraphics[width=4cm]{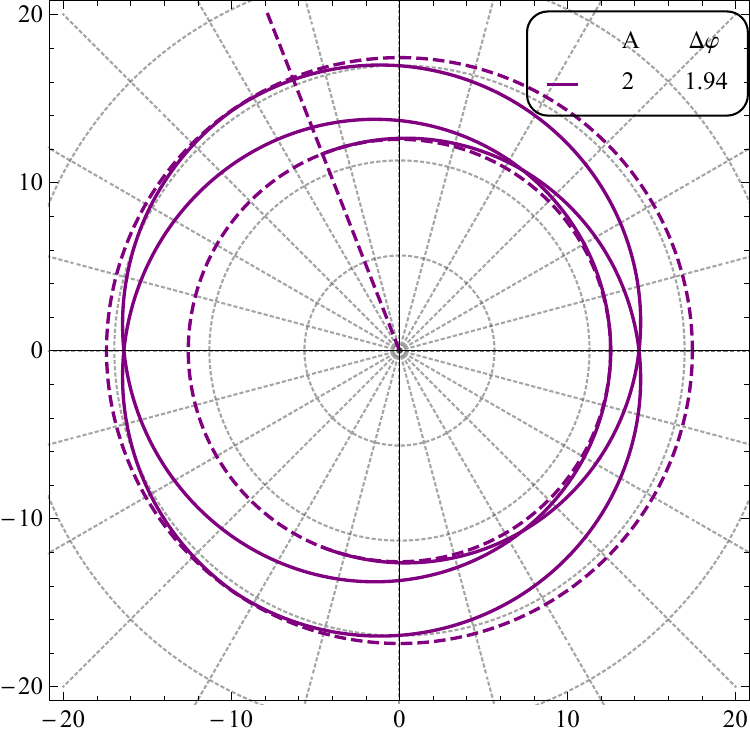}
        \includegraphics[width=4cm]{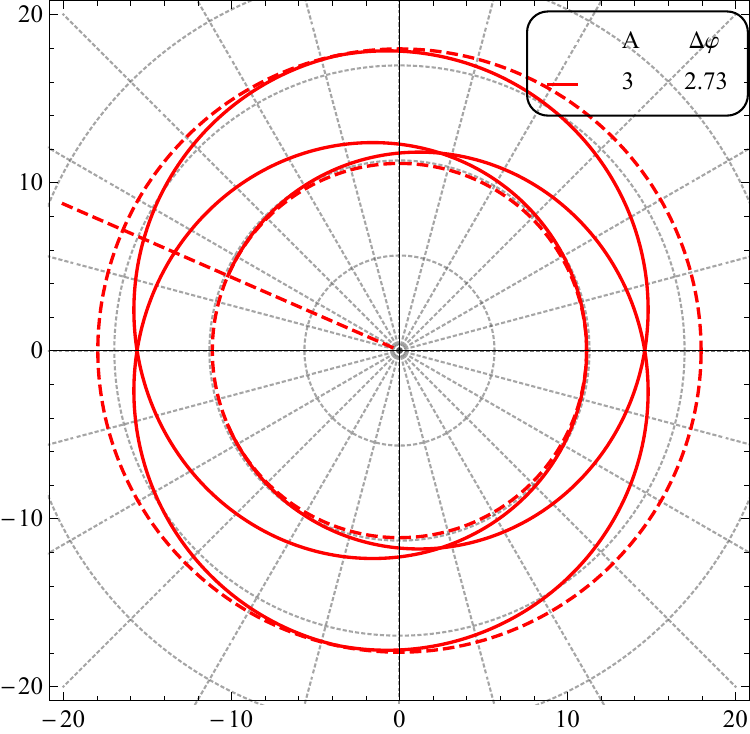}
        \includegraphics[width=4cm]{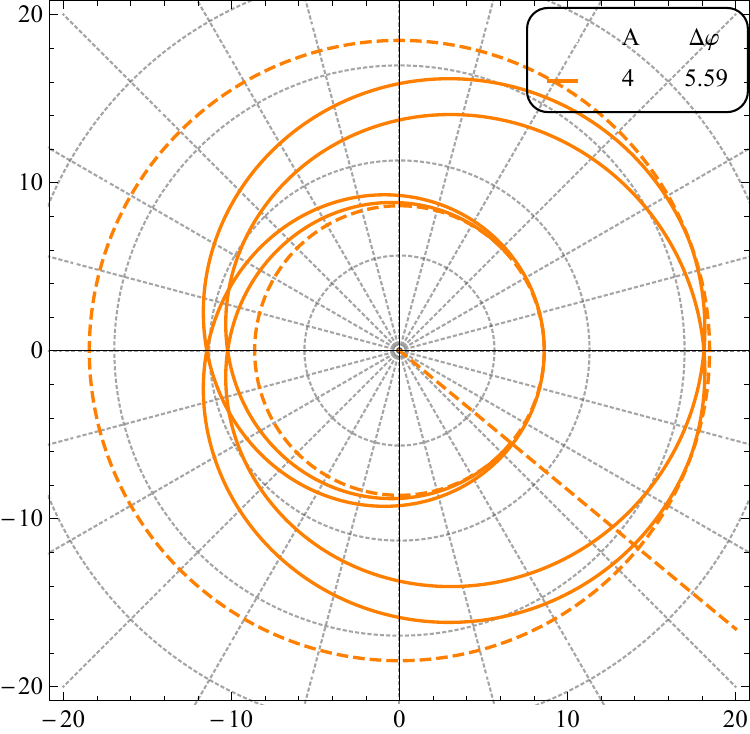}
	\end{center}
	\caption{Planetary orbit $r_P(\phi)$ for different values of the scalar charge $A$.
    Here, $m=1$, $L=4.35$ and $E^2=0.94$. The dashed circles correspond to the perihelion and aphelion distances.}
	\label{figorbitasplanetarias}
\end{figure}

Integrating Eq.~(\ref{w.14}) allows one to obtain the orbital trajectories as well
as the corresponding precession angle $\Delta\varphi$, defined as $\Delta \varphi:= 2\varphi_{(A\to P)}-2\pi$,
where $\varphi_{(A\to P)}$ is the angle from the apoastro to the periastro. In order to visualize the influence of the scalar charge on the
orbital precession, Fig.~\ref{figorbitasplanetarias} displays the planetary
orbits for different values of $A$, while the angular momentum and the energy
are kept fixed. As the scalar charge increases, the precession of the orbit becomes more pronounced.

{\bf{Perihelion precession.}} 
We explicitly assume that the vacuum solution employed provides an accurate description of the exterior gravitational field of a spherically symmetric source such as the Sun, in analogy with the Schwarzschild solution in GR \cite{Chandrasekhar:579245}. This assumption is expected to remain valid in the weak-field regime, where deviations from GR are small. To compute the perihelion shift in a simple and transparent way, we first consider the regime $A<<r$ (\ref{bflat1}) and establish a connection with Kepler's classical problem. Thus, starting from Eq.~\eqref{w.14}, and performing the standard change of variables $u=1/r$, we obtain
\begin{equation}
\label{Binet0}
(u')^2=\frac{2m}{L^2}u-u^2+2m\,u^3+\frac{18mA^2}{5L^2}u^3-A^2u^4-\frac{1-E^2}{L^2}\,,
\end{equation}
where $u'=du/d\varphi$. Now, we convert the first-order Eq.~\eqref{Binet0}  into a second-order  equation by differentiation with respect to $\varphi$, and setting the common factor $u'$, we find
\begin{equation}
\label{Binet}
u''+u=\frac{m}{L^2}+3mu^2+\frac{27mA^2}{5L^2}u^2 -2A^2u^3\,.
\end{equation}
This is the Binet equation 
and exhibits a structure analogous to that of the classical Kepler problem. Now, the following treatment, performed by Adler, Bazin, and Schiffer \cite{Adler},
allows us to derive the formula for the advance of the
perihelia of planetary orbits.
The starting point is to
consider the Binet equation ~\eqref{Binet}. By defining $\tilde{a}=m/L^2$ and introducing the small dimensionless quantity $\epsilon=3m\tilde{a}$, Eq. (\ref{Binet}) can be rewritten as

\begin{equation}
\label{Binet2}
u''+u=\tilde{a}+\frac{\epsilon}{\tilde{a}}u^2+\frac{9\epsilon A^2}{5m}u^2-\frac{2A^2\epsilon}{3m\tilde{a}}u^3\,.
\end{equation}
This can be solved by assuming the solution ansatz $u(\varphi)\approx u_0(\varphi)+\epsilon v(\varphi)+\mathcal{O}(\epsilon^2)$. So, substituting in the differential equation (\ref{Binet2}), we obtain
\begin{equation}
\label{Binet3}
u_0''+u_0+\epsilon(v''+v)=\tilde{a}+\frac{\epsilon}{\tilde{a}}u_0^2+\frac{9\epsilon A^2}{5m}u_0^2-\frac{2A^2\epsilon}{3m\tilde{a}}u_0^3+\mathcal{O}(\epsilon^2)\,.
\end{equation}
Equating the zeroth-order terms in $\epsilon$, yields
\begin{equation}
u_0''+u_0=\tilde{a}\,.
\end{equation}
The solution for $u_0$ is 
\begin{equation}
\label{B0}
u_0=\tilde{a}+\tilde{b}\cos(\varphi+\delta)\,=\frac{1}{\ell}+\frac{e}{\ell}\cos\varphi\,,
\end{equation}
where appropriate axes orientations have been performed given $\delta=0$,  $\tilde{a}=1/\ell$, and $\tilde{b}=e/\ell$, being $\ell$ the latus rectum and $e$ the eccentricity of the Keplerian ellipse. 

Now, equating the first-order $\epsilon$ terms in Eq. (\ref{Binet3}), yields

\begin{equation}
\label{v}
v''+v= \frac{u_0^2}{\tilde{a}}+\frac{9 A^2}{5m}u_0^2-\frac{2A^2}{3m\tilde{a}}u_0^3\,.
\end{equation}
Thus, substituting Eq. (\ref{B0}) in Eq. (\ref{v}), we obtain  
\begin{equation} 
\label{m0}
v''+v=\bar{A}+\bar{B}\cos\varphi+\bar{C}\cos^2\varphi+\bar{D}\cos^3\varphi\,,
\end{equation}
where
\begin{equation}
\bar{A}=\frac{1}{\ell}+\frac{17A^2}{15m\ell^2}\,,\,\, \bar{B}=2e\left(  \frac{1}{\ell}+\frac{4A^2}{5m\ell^2}\right) \,,
\end{equation}
\begin{equation}
\bar{C}=e^2\left(  \frac{1}{\ell}-\frac{A^2}{5m\ell^2}\right)\,,\quad \bar{D}= -\frac{2A^2  e^3}{3m \ell^2}\,.
\end{equation}
Now, using the trigonometric identities $2\cos^2\varphi=1+\cos2\varphi$, and $4\cos^3\varphi=3+\cos 3\varphi$ allows us to write Eq. (\ref{m0}) as 
\begin{equation}  
\label{m1}
v''+v=\tilde{A}+\tilde{B}\cos\varphi+\tilde{C}\cos 2\varphi+\tilde{D}\cos 3\varphi\,,
\end{equation}
where
\begin{equation}
\tilde{A}=\bar{A}+\frac{\bar{C}}{2}\,,\,\, \tilde{B}= \bar{B}+\frac{3\bar{D}}{4}\,,
\end{equation}
\begin{equation}
\tilde{C}=\frac{\bar{C}}{2}\,,\,\, \tilde{D}=\frac{\bar{D}}{4}\,.
\end{equation}
The solution to Eq. (\ref{m1}) can be written as $v=v_A+v_B+v_C+v_D$ with
\begin{equation}
v_A''+v_A=\tilde{A}\,,\,\, v_B''+v_B=\tilde{B}\cos\varphi\,,
\end{equation}
\begin{equation}
v_C''+v_C=\tilde{C}\cos2\varphi\,,\,\, v_D''+v_D=\tilde{D}\cos3\varphi\,,
\end{equation}
whose solutions are 
\begin{equation}
v_A=\tilde{A}\,,\,\, v_B=\frac{\tilde{B}}{2}\varphi\sin\varphi\,,
\end{equation}
\begin{equation}
v_C=-\frac{\tilde{C}}{3}\cos2\varphi\,,\,\, v_D=-\frac{\tilde{D}}{8}\cos3\varphi\,.
\end{equation}
Therefore, the solution is 
\begin{eqnarray}
\notag        u&=&\frac{1}{\ell}+\frac{e}{\ell}\cos\varphi+\epsilon\frac{\tilde{B}}{2}\varphi\sin\varphi+\\
&& +\epsilon\left(\tilde{A}-\frac{\tilde{C}}{3}\cos2\varphi-\frac{\tilde{D}}{8}\cos 3\varphi\right)\,.
\end{eqnarray}
Now, using the following trigonometric identity
$\cos(\varphi-\tilde{\epsilon}\varphi)\approx \cos(\varphi)+\tilde{\epsilon}\varphi \sin(\varphi)$, allows rewriting the solution as
\begin{equation}
\label{uff}
u=\frac{1}{\ell}+\frac{e}{\ell}\cos(\varphi-\tilde{\epsilon}\varphi)+\frac{3m}{\ell}\left(\tilde{A}-\frac{\tilde{C}}{3}\cos2\varphi-\frac{\tilde{D}}{8}\cos 3\varphi\right)\,,
\end{equation}
where
\begin{equation}
\tilde{\epsilon}=\frac{\epsilon\tilde{B}\ell }{2e}=\frac{3m}{\ell}+\frac{12 A^2}{5\ell^2}+\mathcal{O}(e^2)\,.
\end{equation}

In this form, the influence of the various terms on the orbit becomes clear. The fundamental elliptical orbit is described by Eq. (\ref{B0}). The last term introduces small periodic variations in the radial distance of the planet. These variations are subtle and difficult to detect and, because they are periodic, do not affect the precession of the perihelion. However, the parameter $\tilde{\epsilon}\varphi$
in the cosine argument introduces a non-periodic component. Since $\varphi$	
can become significant, its effect is not negligible. Thus, Eq. (\ref{uff}) can be written in the form
\begin{equation}
\label{uf}
u=\frac{1}{\ell}+\frac{e}{\ell}\cos(\varphi-\tilde{\epsilon}\varphi)+\left(\text{periodic terms of order $\epsilon$}\right)\,.
\end{equation}
The perihelion of a planet occurs when $r$ is a minimum or when $u=1/r$  is a maximum.  From Eq. (\ref{uf}) we see that $u$ is maximum when
\begin{eqnarray}
\varphi(1-\tilde{\epsilon})=2\pi n\,,
\end{eqnarray}
or approximately
\begin{eqnarray}
\varphi = 2\pi n(1+\tilde{\epsilon})=:\varphi_n\,.
\end{eqnarray}
Therefore, successive perihelia will occur at intervals of $\Delta\tilde{\varphi}=\varphi_{n+1}-\varphi_{n}= 2\pi (1+\tilde{\epsilon})= 2\pi+\Delta\varphi$,
where $\Delta\varphi=2\pi\tilde{\epsilon}$ and it is equal to
\begin{equation}
\Delta \varphi=\frac{6\pi m}{\ell}+\frac{24\pi A^2}{5\,\ell^2}\,,
\end{equation}
which corresponds to the standard general relativistic contribution, given by the first term, plus an additional correction induced by the regularization of spacetime through the scalar charge $A$. This extra term arises from deviations of the metric from the Schwarzschild geometry due to the presence of the phantom scalar field, and vanishes smoothly in the limit $A \to 0$, where the Schwarzschild result is recovered. 
To apply the above relation to the Solar System, physical dimensions
are restored and the mass parameter is specialized to the geometrized
solar mass,
$
m=M_{\odot}\equiv\frac{GM_{\odot}^{\rm phys}}{c^2}
\simeq 1476.1\,{\rm m}.
$
The resulting perihelion advance can then be expressed in
arcseconds per century
\begin{eqnarray}
\Delta\tilde{\varphi} &=& 100{3600\cdot180\over \pi}f_{rev}\left( \frac{6\pi M_{\astrosun}}{\ell}+\frac{24\pi A^2}{5\,\ell^2} \right)\,,\\
&=&3888\times10^{5}f_{rev}\left( \frac{ M_{\astrosun}}{\ell}+\frac{4 A^2}{5\,\ell^2} \right)\,,
\label{pp}
\end{eqnarray}
in which $f_{rev}$, the revolution frequency, corresponds to the number of orbits per year.
Considering Eq. (\ref{pp}) we constrain the scalar charge $A$,
for Mercury, Venus and Earth.
In Table \ref{tab:horndeski_gamma}, the allowed interval for the scalar charge $A$ is
obtained by taking into account that the correction induced by $A$
in Eq.~(\ref{pp}) is positive and proportional to $A^2$. The lower limit
corresponds to $A=0$, for which the Schwarzschild prediction is
recovered. This value is reported in the fifth column of Table \ref{tab:horndeski_gamma} as
the General Relativity (GR) perihelion shift. The upper limit $A_{\rm max}$ is obtained
by requiring the theoretical perihelion advance to coincide with the
upper edge of the observational interval reported in the second
column, namely $\Delta\phi_{\rm th}(A_{\rm max})
=
\Delta\phi_{\rm obs}+\sigma$. Accordingly, the values displayed in the last column represent the
allowed intervals $0\leq A\leq A_{\rm max}$. For Mercury, Venus,
and Earth we obtain, respectively,
$A_{\rm max}=1.048\times10^{6}\,{\rm m}$,
$1.028\times10^{7}\,{\rm m}$, and
$1.294\times10^{7}\,{\rm m}$. Therefore, among the three planets
considered here, Mercury provides the strongest constraint on the
scalar charge. It is important to distinguish the normalization used in the
strong-field numerical analysis from the dimensional Solar-System
application. Throughout the numerical geodesic analysis we set
$m=1$, so that $r$, $A$, and $L$ are measured in units of the mass
parameter $m$. For the Solar-System test, physical dimensions are
restored and the bounds on $A$ reported in Table~\ref{tab:horndeski_gamma}
are given in meters.

\begin{table}[htbp]
\centering
\renewcommand{\arraystretch}{1.2}

\resizebox{\columnwidth}{!}{%
\begin{tabular}{|c|c|c|c|c|c|}
\hline
\text{Planet} &
\begin{tabular}[c]{@{}l@{}}
Observed \\
shift. \\
Seconds \\
per century
\end{tabular} &
\begin{tabular}[c]{@{}l@{}}
Latus \\
rectum $(\ell)$ \\
$(\times 10^{10}\ \rm m)$
\end{tabular} &
\begin{tabular}[c]{@{}l@{}}
Revolution \\
frequency \\
$(f_{\rm{rev}}).$ \\
Orbits per \\
annum
\end{tabular} &
\begin{tabular}[c]{@{}c@{}}
GR shift. \\
Seconds \\
per century\\
\end{tabular} &
\begin{tabular}[c]{@{}c@{}}
A (m)\\
\end{tabular} \\
\hline
Mercury & $43.11 \pm 0.45$ & 5.53 & 4.15  & 43.10 & $0\leq A\leq 1.048 \times 10^{6}$\\
Venus   & $8.4 \pm 4.8$    & 10.8 & 1.622 & 8.62  & $0\leq A \leq 1.028 \times 10^{7}$ \\
Earth   & $5 \pm 1.2$      & 14.9 & 1     & 3.85 & $0\leq A \leq 1.294 \times 10^{7}$ \\
\hline
\end{tabular}%
}

\caption{Constraints on the scalar charge $A$ from Solar System perihelion precession.}
\label{tab:horndeski_gamma}
\end{table}

\subsubsection{Second kind trajectories} 

Second kind trajectories correspond to bounded configurations in which the particle starts at a finite radial distance and inevitably plunges into the black hole after reaching a single turning point. In this regime, equation $E^{2}-V^{2}(r)=0$
admits a single positive root, which determines the turning point $r_F$. The particle approaches $r_F$, reverses its motion, and subsequently falls into the event horizon. Fig. \ref{figsecondkind} illustrates representative second kind trajectories for $m=1$ and different values of the scalar charge $A$.  As the scalar charge increases, the position of the turning point $r_F$--and $R_F$--shifts outward, indicating that the scalar hair modifies the radial structure of the spacetime and alters the capture dynamics. In the limit $A\to0$, the effective potential reduces to the Schwarzschild case, and the second kind trajectories recover the standard plunging behavior of General Relativity.

\begin{figure}[H]
	\begin{center}
		\includegraphics[width=6.5cm]{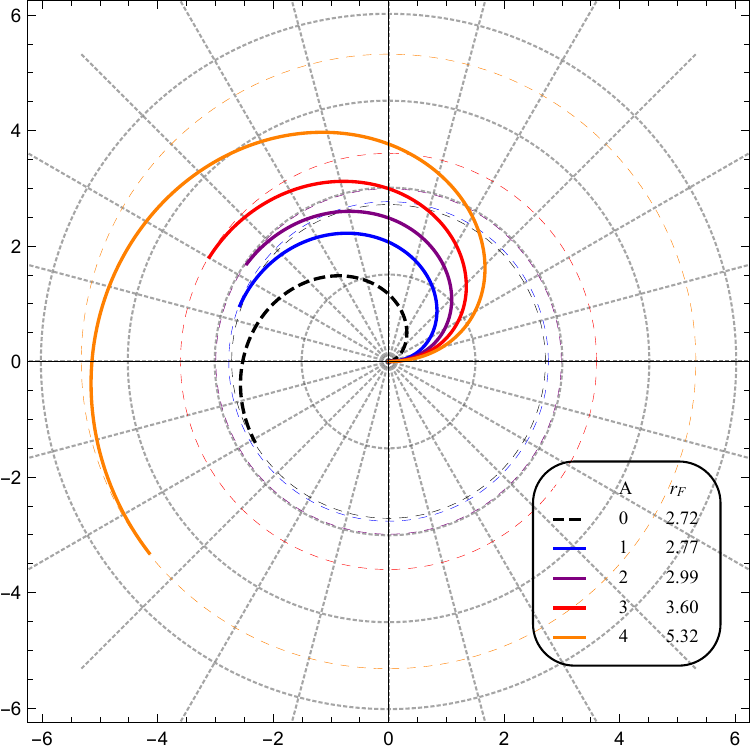}
	\end{center}
	\caption{Second kind trajectories for a massive particle and for different values of the scalar charge $A$. The dashed circle corresponds to the turning point. Here, $m=1$, $L=4.35$ and $E^2=0.94$.}
	\label{figsecondkind}
\end{figure}

\subsubsection{Critical trajectories with $E<1$}

Using the angular geodesic equation (\ref{w.14}), the orbital motion can be expressed in integral form
\begin{eqnarray} \label{angular4}
\varphi(r) = \pm \int_{R_0}^{r} 
\frac{L\,dr'}{R(r')^{2}\sqrt{E^{2}-V^{2}(r')}}\,,
\end{eqnarray}
where $R_0$ denotes the initial radial position. 
The qualitative behavior of the trajectories is determined by the structure of the radial roots of the equation
\begin{equation}
E^{2}-V^{2}(r)=0,
\end{equation}
which fixes the turning points. 
Critical trajectories 
are controlled by the unstable circular orbit, mathematically, they occur when the radial equation
$E^{2}-V^{2}(r)=0$ develops a double root at $r=r_U$, namely $E=E_U:=V(r_U)$, $V'(r_U)=0$. Under this condition, the radial turning point merges with the unstable circular orbit and the angular integral diverges logarithmically as $r\to r_U$. As a consequence, the particle asymptotically approaches the unstable orbit while accumulating an arbitrarily large azimuthal angle, producing characteristic spiral behavior.

\begin{figure}[H]
	\begin{center}
		\includegraphics[width=6 cm]{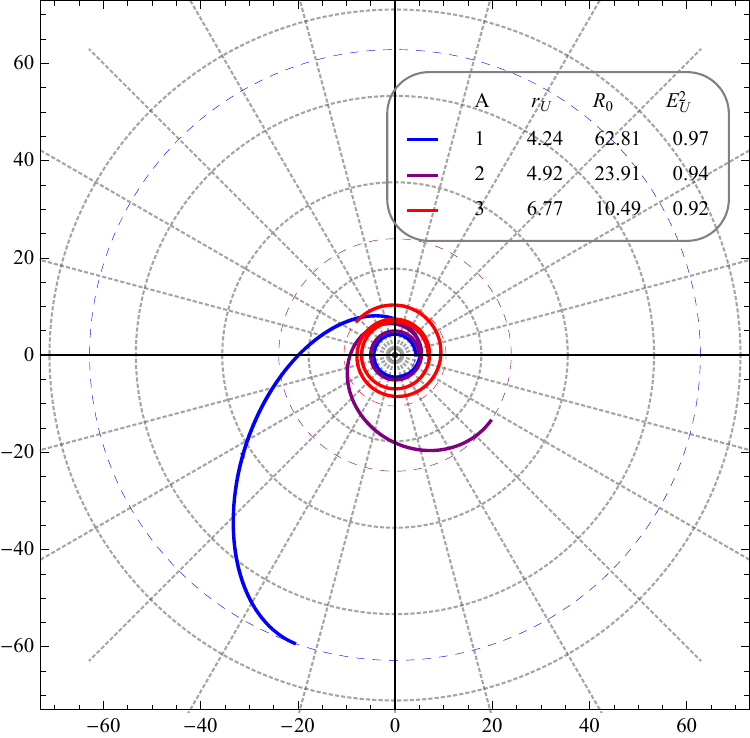}
		\includegraphics[width=6 cm]{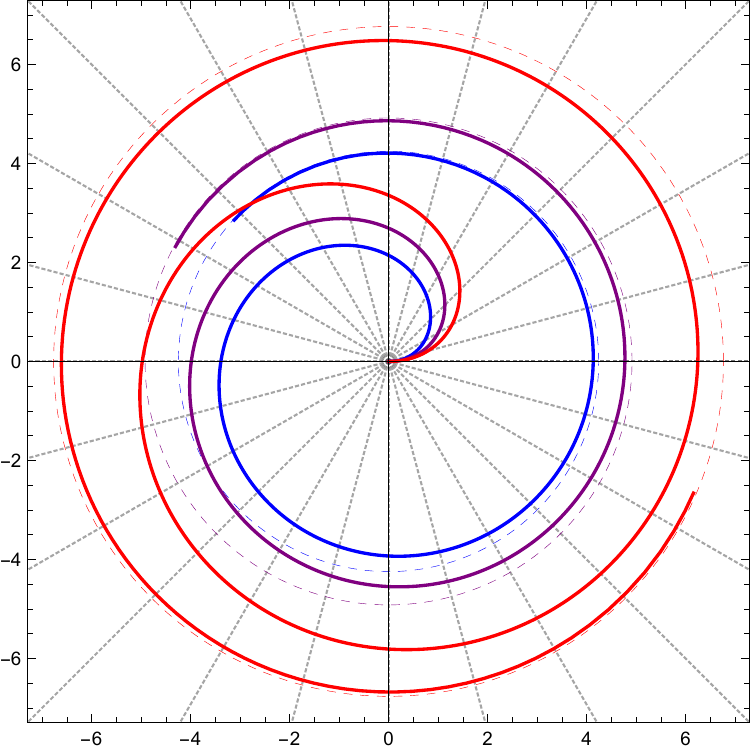}
	\end{center}
	\caption{ Critical trajectories with  $m=1$,  and $L=3.95$ for different values of the scalar charge $A$.  CFK trajectories (top panel) and CSK trajectories (bottom panel) with $E_U<1$. The dashed circle represents $R_0$ in the top panel and $r_U$ in the bottom panel.   }
	\label{figcriticalbounded1}
\end{figure}

Fig. \ref{figcriticalbounded1} illustrates the CFK and CSK trajectories for $m=1$ and fixed angular momentum $L=3.95$, in the regime $E_U<1$. These critical orbits represent the separatrix between long-lived bound motion and plunging trajectories. As the scalar charge $A$ increases, the unstable orbit $r_U$--and $R_U$-- shifts outward and the associated critical energy decreases.
In the limit $A\to0$, the separatrix structure smoothly reduces to the Schwarzschild case.

\begin{figure}[!h]
	\begin{center}
\includegraphics[width=65mm]{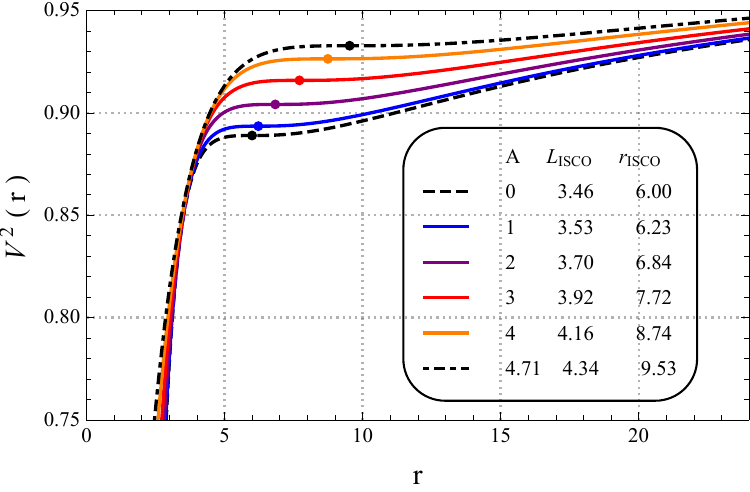}
	\end{center}
	\caption{Effective potential $V^2(r)$ for different values of the scalar charge $A$. Here, $m=1$, and $L=L_{\rm{ISCO}}$. The dot corresponds to $r_{\rm{ISCO}}$.}
	\label{potentialisco}
\end{figure}

A special critical configuration occurs when the local maximum and minimum of the effective potential merge into a single inflection point. This corresponds to the condition
\begin{equation}
V'(r_{\rm ISCO})=0,
\qquad
V''(r_{\rm ISCO})=0,
\end{equation}
which defines the innermost stable circular orbit (ISCO).
Fig.~\ref{potentialisco} shows the behavior of the effective potential for different values of the scalar charge $A$ at $L=L_{\rm ISCO}$. 
As $A$ increases, the angular momentum $L_{\rm ISCO}$ increases
and the ISCO moves outward. Importantly, this outward displacement
is observed both in the coordinate radius $r_{\rm ISCO}$ and in the
invariant areal radius $R_{\rm ISCO}$. For instance,
$r_{\rm ISCO}$ increases from $6.00$ to $9.53$ as $A$ varies from
$0$ to $4.71$, while the corresponding areal radius increases from
$R_{\rm ISCO}=6.00$ to $R_{\rm ISCO}=10.63$. The complete values
are reported in Table~\ref{tab:characteristic_radii}. Hence, the
outward displacement of the ISCO represents a genuine modification
of the characteristic geometric scale rather than a consequence of
the radial-coordinate choice.

\begin{figure}[H]
	\begin{center}
		\includegraphics[width=6. cm]{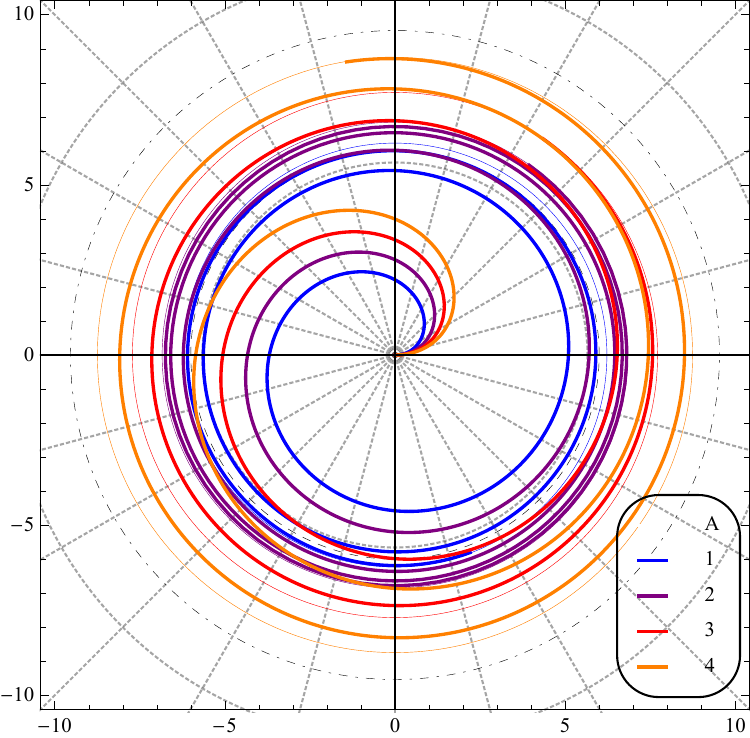}
	\end{center}
	\caption{ Critical trajectories with  $m=1$  and $L_{\rm{ISCO}}$ for different values of the scalar charge $A$. The larger circle represents $r_{\text{ISCO}}$.}
	\label{figcriticalISCO}
\end{figure}

The corresponding critical trajectories are shown in Fig.~\ref{figcriticalISCO}. 
As $A$ increases, the ISCO moves outward and the critical trajectory adjusts accordingly.
In the limit $A\to0$, the inflection condition reduces to the Schwarzschild case, recovering the standard result $r_{\rm ISCO}=6m$ and the corresponding separatrix structure.

\begin{figure}[!h]
	\begin{center}
\includegraphics[width=65mm]{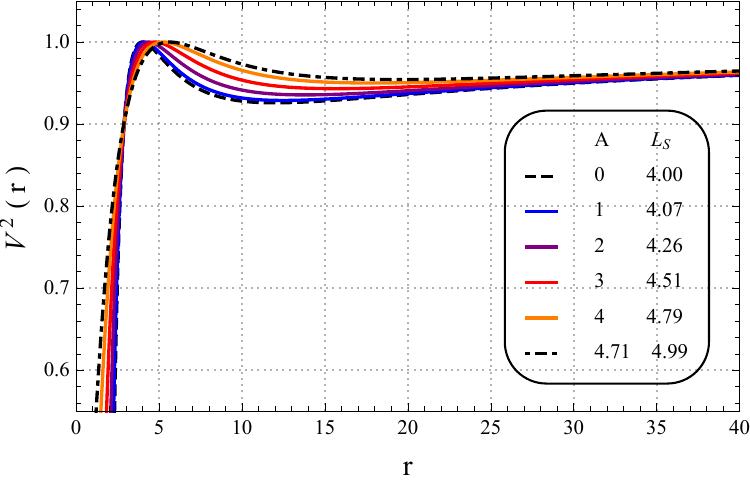}
\includegraphics[width=65mm]{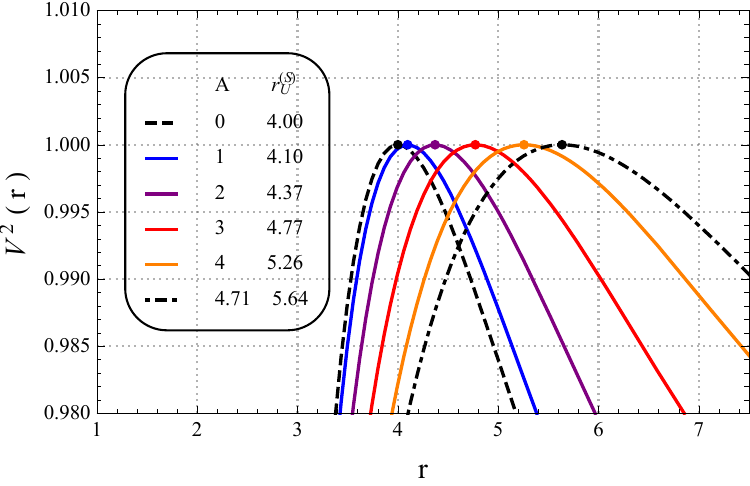}
	\end{center}
	\caption{Effective potential $V^2(r)$ for different values of the parameter $A$, with $m=1$ and $E=1$. The upper panel shows the global structure of the potential, whereas the lower panel zooms into the vicinity of the maximum, corresponding to the unstable circular orbit $r_u$.
    }
	\label{PotLS}
\end{figure}

The onset of the unbounded regime is illustrated in Fig.~\ref{PotLS}, where the
behavior of the effective potential is displayed. For each value of the
scalar charge $A$, the angular momentum $L=L_S$ is chosen such that the
maximum of the effective potential satisfies $V^2(r_U)=1$, where $r_U$ denotes the unstable circular orbit. In this way, the value
of $L_S$ corresponds to the threshold that separates capture from scattering
trajectories.
Therefore, the maximum of the effective potential coincides with its asymptotic value. 
Fig. \ref{PotLS} shows that the critical angular momentum $L_S$ increases
with the scalar charge $A$. The unstable circular orbit defining
this threshold also moves outward. To distinguish it from the
unstable circular orbit considered at fixed $L$ in Fig.~\ref{PotLS} (lower panel), we denote
its radius by $r_U^{(S)}$ and the corresponding areal radius by
$R_U^{(S)}$. Both quantities increase monotonically with $A$, as
shown in Table~\ref{tab:characteristic_radii}. In particular,
$r_U^{(S)}$ changes from $4.00$ in the Schwarzschild limit to
$5.64$ near the critical scalar charge, while
$R_U^{(S)}$ changes from $4.00$ to $7.35$. Thus, the displacement
of the capture--scattering threshold toward larger radial scales
persists when expressed in terms of the invariant areal radius.

On the other hand, when $L>L_S$, the maximum of the potential lies above unity, $E_U>1$, and a genuine scattering zone appears. Conversely, for $L<L_S$, no scattering is possible and all trajectories with $E\ge1$ inevitably plunge into the horizon.\\

We emphasize that the curves corresponding to $A\simeq4.71$
represent the limiting critical configuration of the black-hole
branch, for which the event horizon coincides with the throat at $r=0$. The notions of plunge, capture, and capture--scattering
separatrix employed here rely on the presence of an event horizon
and therefore should not be directly extended to
$A>A_{\rm crit}$. In this regime, the horizon moves to the $r<0$ side of the extended geometry, while the $r>0$ region connected to the asymptotically flat exterior is horizonless and contains the throat at $r=0$.\\

Table~\ref{tab:characteristic_radii} also highlights the importance
of distinguishing coordinate and invariant radial scales. The most
striking example is the event horizon. As $A$ increases, its
coordinate position $r_+$ moves inward and approaches zero at the
critical configuration, whereas its invariant areal radius $R_+$
increases and remains finite. Therefore, the decrease of $r_+$ does
not correspond to a shrinking physical horizon. In contrast, the
unstable circular orbit, the ISCO, and the unstable orbit defining
the scattering threshold move outward in both $r$ and $R$. The
stable circular orbit considered at fixed $L=4.35$ moves inward in
both radial descriptions. These results show that the changes in the
characteristic orbital scales produced by the scalar charge are
genuine geometric effects, although their numerical magnitude
depends on the radial variable used to describe them.

\begin{table}[htbp]
\centering
\renewcommand{\arraystretch}{1.2}
\resizebox{\columnwidth}{!}{%
\begin{tabular}{c|cc|cc|cc|cc|cc}
\hline\hline
$A$
& $r_+$ & $R_+$
& $r_U$ & $R_U$
& $r_S$ & $R_S$
& $r_{\rm ISCO}$ & $R_{\rm ISCO}$
& $r_U^{(S)}$ & $R_U^{(S)}$ \\
\hline
0
& 2.000 & 2.000
& 3.74 & 3.74
& 15.18 & 15.18
& 6.00 & 6.00
& 4.00 & 4.00 \\

1
& 1.901 & 2.148
& 3.86 & 3.99
& 15.06 & 15.09
& 6.23 & 6.31
& 4.10 & 4.22 \\

2
& 1.610 & 2.568
& 4.27 & 4.72
& 14.67 & 14.81
& 6.84 & 7.13
& 4.37 & 4.81 \\

3
& 1.146 & 3.211
& 5.05 & 5.87
& 13.94 & 14.26
& 7.72 & 8.28
& 4.77 & 5.64 \\

4
& 0.526 & 4.034
& 6.45 & 7.59
& 12.62 & 13.24
& 8.74 & 9.61
& 5.26 & 6.61 \\

4.71
& 0.000 & 4.710
& 8.87 & 10.04
& 10.27 & 11.30
& 9.53 & 10.63
& 5.64 & 7.35 \\
\hline\hline
\end{tabular}}
\caption{Coordinate and invariant areal radii of the main
characteristic orbits for $m=1$. The areal radius is
$R(r)=\sqrt{r^2+A^2}$. The quantities $r_U$ and $r_S$ refer to the
unstable and stable circular orbits shown in Fig.~2 for $L=4.35$,
whereas $r_U^{(S)}$ denotes the unstable circular orbit at the
scattering threshold, defined by $V^2(r_U^{(S)})=1$ and
$L=L_S$.}
\label{tab:characteristic_radii}
\end{table}

These trends also indicate a progressive rearrangement of the
characteristic orbital scales as $A$ approaches $A_{\rm crit}$:
the unstable circular orbit and the ISCO move outward, while the
stable circular orbit considered at fixed $L$ moves inward. The
consequences of this behavior for unbounded trajectories are discussed
below.

\subsection{Unbounded Orbits}

As we mentioned, when $L>L_S$, the effective potential develops a local maximum that satisfies $E_U>1$, giving rise to a genuine scattering region. In this case, particles with $1 \le E < E_U$ are reflected by the potential barrier, whereas particles with $E>E_U$ overcome it and fall into the black hole. The unstable circular orbit at $r=r_U$ defines the separatrix between these two behaviors.

\begin{figure}[!h]
	\begin{center}
\includegraphics[width=65mm]{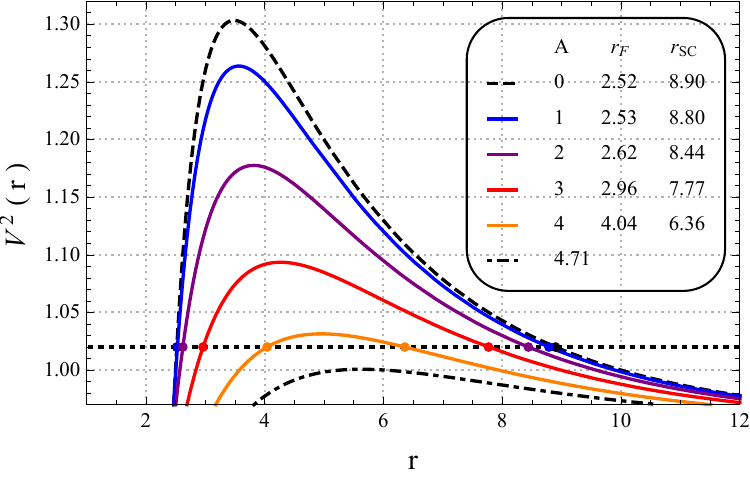}
	\end{center}
	\caption{Effective potential $V^2(r)$ for different values of the parameter $A$, with $m=1$ and $L=5$. The horizontal line represents the energy level $E^2=1.02$. The intersections define the return distance $r_F$, while $r_{SC}$ denotes the scattering distance.
   }
	\label{potentials}
\end{figure}

Fig. \ref{potentials} illustrates the behavior of the effective potential for fixed
parameters $m=1$, $L=5$, and $E^2=1.02$, considering several values
of the scalar charge $A$. The horizontal line represents the energy
level $E^2$, and its intersections with the effective potential
determine the radial turning points. The inner intersection $r_F$
corresponds to the turning point associated with second-kind
trajectories, whereas the outer intersection $r_{\rm SC}$ determines
the distance of closest approach for scattering trajectories.

As the scalar charge increases, the two turning points evolve in
opposite directions. In terms of the coordinate radius, $r_F$
increases from $2.52$ for $A=0$ to $4.04$ for $A=4$, whereas
$r_{\rm SC}$ decreases from $8.90$ to $6.36$. Importantly, the same
qualitative behavior is obtained in terms of the invariant areal
radius. The corresponding inner turning point increases from
$R_F=2.52$ to $R_F=5.69$, while the scattering radius decreases from
$R_{\rm SC}=8.90$ to $R_{\rm SC}=7.51$. The complete values are
reported in Table~\ref{tab:scattering_radii}. Therefore, the
outward displacement of the inner turning point and the inward
displacement of the scattering distance persist when described using
the invariant geometric radial scale.

For the limiting configuration $A\simeq4.71$, no turning points are
present for the fixed energy $E^2=1.02$. In this case, the maximum
of the effective potential satisfies $E_U^2\simeq1.0007<E^2$, so
that the particle overcomes the potential barrier and belongs to the
capture regime. Consequently, $r_F$, $R_F$, $r_{\rm SC}$, and
$R_{\rm SC}$ are not defined for this particular choice of the
particle energy.

\begin{table}[htbp]
\centering
\renewcommand{\arraystretch}{1.2}
\begin{tabular}{c|cc|cc}
\hline\hline
$A$
& $r_F$ & $R_F$
& $r_{\rm SC}$ & $R_{\rm SC}$ \\
\hline
0    & 2.52 & 2.52 & 8.90 & 8.90 \\
1    & 2.53 & 2.72 & 8.80 & 8.86 \\
2    & 2.62 & 3.30 & 8.44 & 8.67 \\
3    & 2.96 & 4.21 & 7.77 & 8.33 \\
4    & 4.04 & 5.69 & 6.36 & 7.51 \\
4.71 & --   & --   & --   & --   \\
\hline\hline
\end{tabular}
\caption{Coordinate and invariant areal radii of the turning points
shown in Fig.~\ref{potentials} for $m=1$, $L=5$, and $E^2=1.02$.
Here, $r_F$ denotes the inner turning point associated with
second-kind trajectories and $r_{\rm SC}$ the distance of closest
approach for scattering trajectories, while
$R_F=\sqrt{r_F^2+A^2}$ and
$R_{\rm SC}=\sqrt{r_{\rm SC}^2+A^2}$ are their corresponding
invariant areal radii. For $A\simeq4.71$, no turning points exist
for this fixed energy because the particle energy lies above the
maximum of the effective potential.}
\label{tab:scattering_radii}
\end{table}

The behavior displayed in Fig.~\ref{potentials} and Table~\ref{tab:scattering_radii} provides a useful
illustration of how the geodesic structure changes as the scalar
charge approaches its critical value from the black-hole branch.
As $A\to A_{\rm crit}^{-}$, the effective-potential barrier becomes
progressively lower for the fixed angular momentum considered here.
Accordingly, the inner turning point moves outward, whereas the
scattering distance moves inward, reducing the radial region
associated with reflected trajectories. The same qualitative behavior
is observed in terms of the invariant areal radii $R_F$ and
$R_{\rm SC}$.

At the same time, the critical energy associated with the maximum of
the effective potential decreases toward its asymptotic value. For
$A\simeq4.71$, we find $E_U^2\simeq1.0007$, so that for the fixed
energy $E^2=1.02$ used in Fig.~10 the particle energy exceeds the
potential barrier. Consequently, the two turning points disappear and
the corresponding trajectory changes from scattering to capture.
Thus, for fixed particle energy and angular momentum, approaching the
critical scalar charge can qualitatively modify the character of the
timelike trajectory.

This interpretation applies to the approach to the critical
configuration from the black-hole branch. For $A>A_{\rm crit}$ the
event horizon is absent from the region $r \geq 0$, and the spacetime enters the
wormhole-like branch. In that regime, the notions of plunge and
capture used here are no longer directly applicable, and the geodesic
structure would require a separate analysis.

\newpage

\subsubsection{Scattering trajectories}
\begin{figure}[H] 
    \begin{center}
        \includegraphics[width=5.0cm]{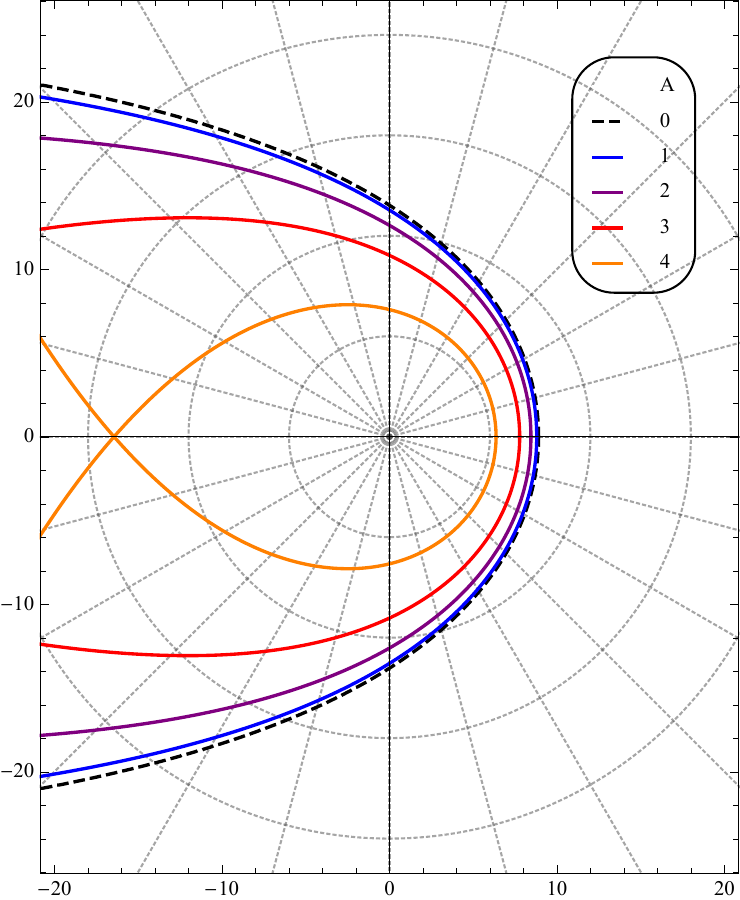}
    \end{center}
    \caption{ Scattering trajectories of a massive particle for different values of the parameter $A$, with $m=1$, $L=5$ and $E^2=1.02$.
}
    \label{figScattering}
\end{figure}

Fig.~\ref{figScattering} displays the corresponding scattering trajectories. These trajectories illustrate
the unbounded motion in the regime $L>L_S$, where particles with $1<E^2<E_U^2$ coming from
infinity are scattered by the effective potential barrier and eventually
escape back to infinity. The closest approach point corresponds to the
turning point previously identified as the scattering radius $r_{SC}$.
As the scalar charge increases, for fixed values of $L$ and $E$,
the scattering angle increases, while the distance of closest
approach decreases. This behavior is observed in both the coordinate
radius $r_{\rm SC}$ and the invariant areal radius $R_{\rm SC}$,
confirming that the reduction of the closest-approach scale is a
geometric effect of the scalar hair.

\subsubsection{Second kind trajectories} 

Here, particles located outside the event horizon reach a turning point
$r_F$, see Fig. \ref{potentials}, and subsequently plunge into the black hole. Fig.~\ref{figsecondkind} illustrates representative second kind trajectories for $1<E^2<E_U^2$. As the scalar charge increases, for fixed values of $L$, the turning point $r_F$-- and $R_F$-- increases.

\begin{figure}[H]
	\begin{center}
		\includegraphics[width=6.0cm]{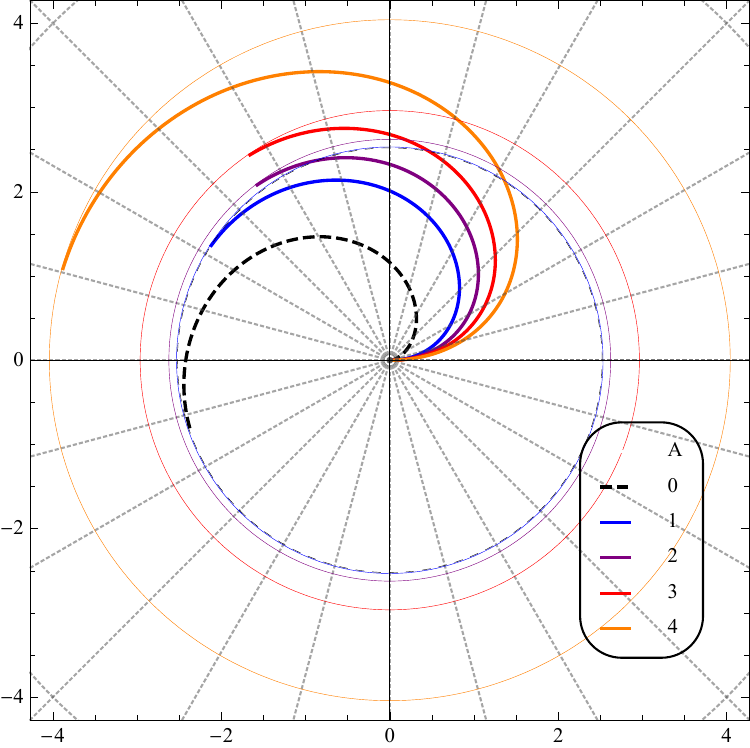}
	\end{center}
	\caption{Second kind trajectories of a massive particle for different values of the parameter $A$. The circles represent the turning point $r_F$. Here, $m=1$, $L=5$ and $E^2=1.02$}
	\label{figsecondkind}
\end{figure}

\subsubsection{Critical trajectories}

\begin{figure}[H]
	\begin{center}
		\includegraphics[width=6cm]{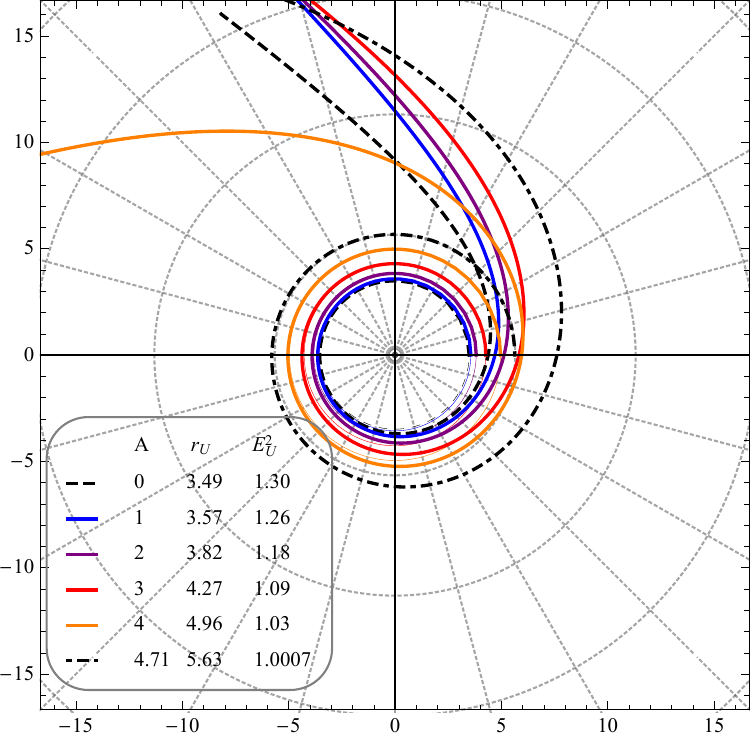}
				\includegraphics[width=6cm]{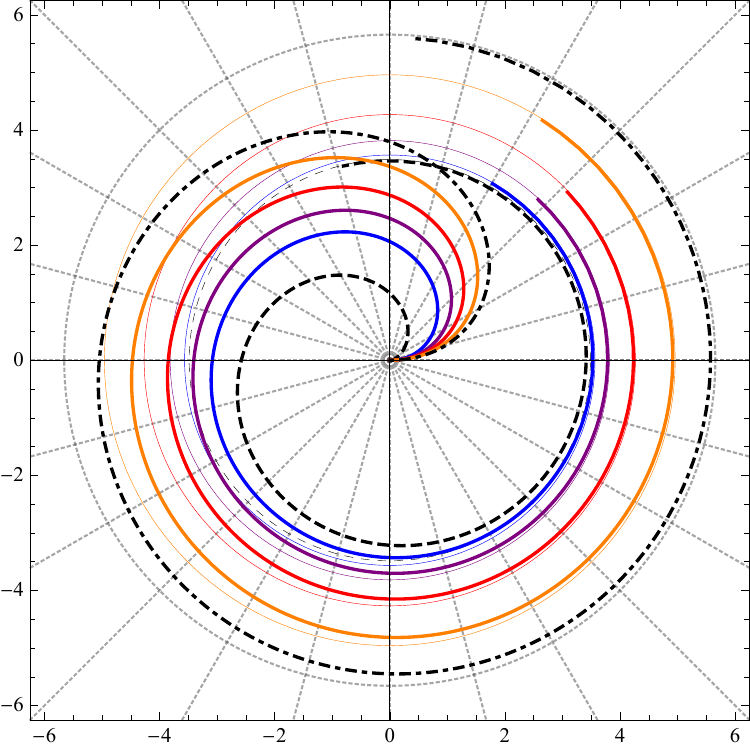}
	\end{center}
	\caption{ Critical trajectories for different values of the parameter $A$, with  $m=1$  and $L=5$. CFK trajectories (top panel) and CSK trajectories (bottom panel) with $E=E_U$.
		Here, the circles indicate the unstable circular radius $r_U$.   }
	\label{figcriticalunbounded}
\end{figure}

In the unbounded regime, a critical configuration occurs when the particle energy coincides with the maximum of the effective potential.
Fig.~\ref{figcriticalunbounded} displays the critical first-kind (CFK) and critical second-kind (CSK) trajectories in this regime. In the CFK case, the particle approaches the unstable circular orbit from the exterior region ($r>r_U$) and asymptotically spirals toward $r_U$ without crossing it (top panel). In the CSK configuration, the particle originates in an interior region ($r_+<r<r_U$) and spirals toward the same unstable orbit from smaller radii (bottom panel).

These critical trajectories represent the boundary between reflection and capture: for $E<E_U$, the motion corresponds to scattering, while for $E>E_U$, the particle inevitably falls into the horizon. The scalar charge $A$ modifies both the position of the unstable circular orbit and the critical energy $E_U$, thus shifting the separatrix. 
In the limit $A\to0$, the standard Schwarzschild critical configuration is recovered.

\subsubsection{Capture zone}

\begin{figure}[H]
	\begin{center}
		\includegraphics[width=60mm]{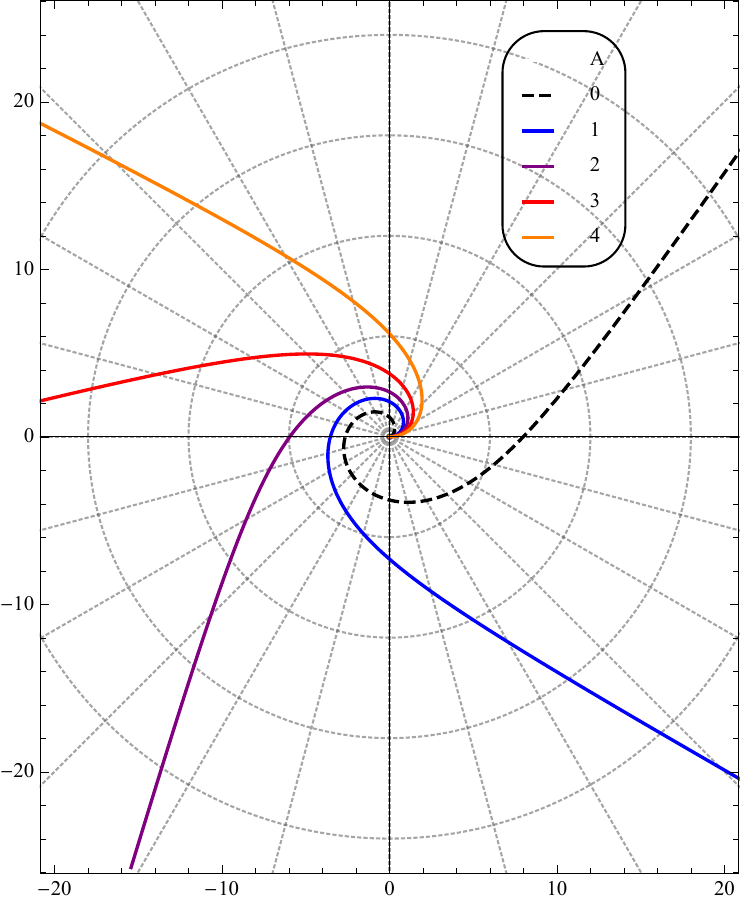}
	\end{center}
	\caption{Capture-zone trajectories for different values of the
scalar charge $A$. Depending on the initial conditions, particles
can plunge into the horizon or escape to infinity. Here,
$m=1$, $L=5$, and $E^2=1.02$.
}
	\label{captura}
\end{figure}

In this regime, the particles satisfy $E>E_U$, corresponding to the
capture zone. In this region the potential barrier can be overcome, and
the particle either plunges into the black hole or escapes to spatial
infinity depending on its trajectory. Representative trajectories are
shown in Fig.~\ref{captura} for several values of the scalar charge $A$. As $A$
increases, the radius of the unstable circular orbit $r_U$-- and $R_U$-- shifts outward,
modifying the trajectories and expanding the region
where the particles are captured.

\section{Motion with $L=0$} 
\label{L0}

\begin{figure}[!h]
	\begin{center}
\includegraphics[width=65mm]{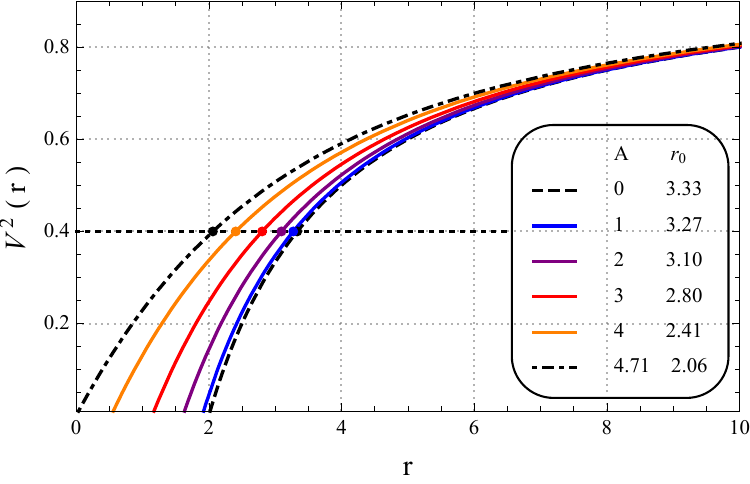}
	\end{center}
	\caption{Effective potential $V^2(r)$ for different values of the scalar charge $A$. Here, $m=1$, $L=0$ and $E^2=0.4$.}
	\label{potentialr}
\end{figure}

In this case, the effective potential reduces to 
$V^2(r)=b(r)$,
so that 
the dynamics is entirely
determined by the metric function $b(r)$. As shown in Fig.~\ref{potentialr}, the potential
vanishes at the event horizon $r=r_+$ and increases monotonically toward
its asymptotic value at large distances.  Although the scalar charge modifies the shape of the potential and the position of the horizon, it does not introduce new qualitative
features in the  dynamics.

The motion of the particles is therefore controlled by the relation between the
energy level $E^2$ and the function $b(r)$. When $E^2<1$, the motion is bounded between the event horizon and a turning point $r_0$, real roots of $E^2=b(r)$. Otherwise, the particle inevitably falls into the black hole. For $E\geq 1$, the particle may escape to spatial infinity or plunge to the horizon, depending on the initial conditions.

The equations governing this type of motion follow directly from Eqs.~(\ref{w.12}) and (\ref{w.13}), yielding

\begin{eqnarray}
\tau(r) &=&\pm \int^{r}_{r_0}{dr'\over \sqrt{E^2-b(r')}}\,, \label{inttau} \\
t(r) &=& \pm \int^{r}_{r_0}{E\,dr'\over b(r') \sqrt{E^2-b(r')}}\,, \label{intt}
\end{eqnarray}

\subsection{Bounded trajectories: $0<E<1$}

\begin{figure}[H]
	\begin{center}
		\includegraphics[width=6.5cm]{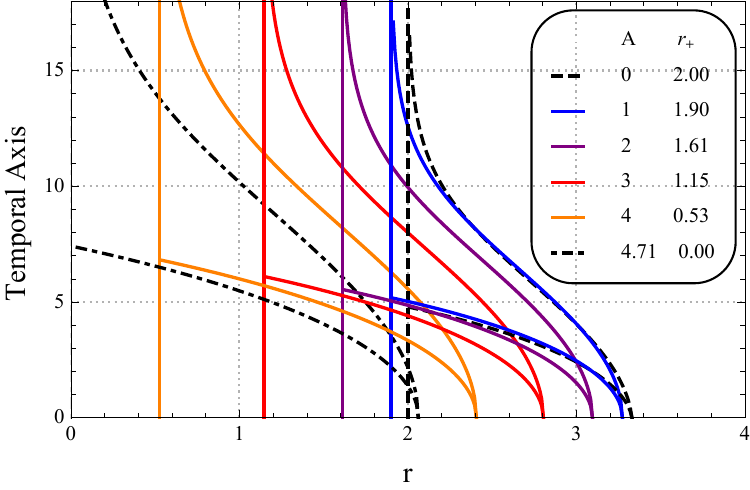}
	\end{center}
	\caption{Proper and coordinate time for the bounded trajectories of test particles with $L=0$ and for different values of the scalar charge $A$ . The vertical lines indicate the corresponding event horizons.
		Here,   $m=1$ and $E^2=0.4$.}
	\label{f2RN}
\end{figure}

The behavior of the proper time $\tau$ and the coordinate time $t$ as functions of the radial coordinate is shown in Fig.~\ref{f2RN}. 
A clear qualitative distinction emerges between the two temporal parameters.
From Eq.~(\ref{inttau}), the proper time required for the particle to reach the event horizon is finite. 
Near $r=r_+$, since $V^2(r_+)=0$, the integrand behaves regularly and the radial motion proceeds smoothly across the horizon in a finite amount of proper time.

In contrast, the coordinate time $t(r)$ exhibits a logarithmic divergence as $r \to r_+$. 
This divergence originates from the factor $b(r)$ that appears in the denominator of Eq.~(\ref{intt}), which vanishes linearly at the horizon. 
Expanding near $r_+$, $b(r) \sim b'(r_+)(r-r_+)$,
one finds
$t(r) \sim - \frac{1}{b'(r_+)} \ln|\frac{r-r_+}{r_0-r_+}|$,
which explains the asymptotic growth shown in Fig.~\ref{f2RN}.

Therefore, while the infalling particle crosses the horizon in finite proper time, an observer located at spatial infinity would assign an infinite coordinate time to this crossing. 
This behavior mirrors the standard Schwarzschild case and confirms that, despite the presence of a scalar charge, the causal structure at the horizon remains unchanged. Notice that as $A$ increases, for a fixed value of the energy $E$, the position of the turning point shifts toward smaller radii $r_0$, see Fig. \ref{potentialr}. 
Consequently, the logarithmic
divergence of the coordinate time occurs at progressively smaller values
of $r$. At the same time, the proper time required for the particle to reach
the horizon slightly increases.

\subsection{Unbounded trajectories: $E \geq 1$}

\begin{figure}[H]
	\begin{center}
		\includegraphics[width=6.5cm]{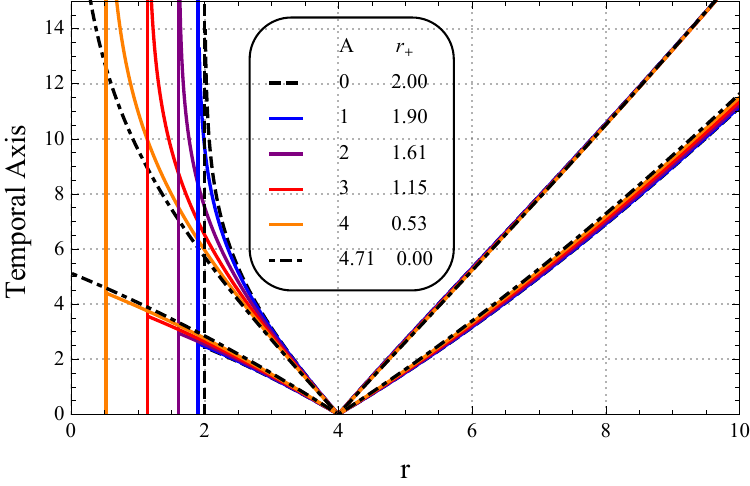}
	\end{center}
	\caption{Proper and coordinate time for the unbounded trajectories of test particles with $L=0$ and for different values of the scalar charge $A$. Here, $m=1$, $r_0=4$ and $E=1$.}
	\label{frub}
\end{figure}

In this regime, the particle possesses sufficient energy to reach spatial infinity if initially directed outward, or to fall from infinity toward the black hole.
For infalling trajectories, the particle approaches the event horizon monotonically.  As shown in Fig. \ref{frub}, the proper time required to reach $r_+$ remains finite, while the coordinate time diverges as $r \to r_+$.

For outgoing motion the particle moves toward spatial infinity.
Both the proper time and the coordinate time increase monotonically
and diverge as $r\rightarrow\infty$.
The scalar charge modifies the evolution through the function $b(r)$, but does not introduce qualitative differences with respect to the Schwarzschild case in the $L=0$ sector.

\section{Conclusions}
\label{FR}

In this work, we have investigated the timelike geodesic structure of asymptotically flat regular black holes supported by a phantom scalar field characterized by a scalar charge $A$. 
This parameter removes the central singularity and continuously deforms the Schwarzschild geometry while preserving asymptotic flatness.

We derived the equations of motion for massive particles and performed a systematic classification of trajectories according to the values of the conserved energy $E$ and angular momentum $L$. 
For $L \neq 0$, both bounded ($E < 1$) and the unbounded ($E \ge 1$) regimes were analyzed, including circular orbits, planetary motion, second kind trajectories, critical configurations, and scattering solutions. 
For the motion with vanishing angular momentum, we studied both bounded and unbounded trajectories in terms of proper and coordinate time.

A central result of our analysis is that the scalar charge $A$ induces a genuine geometric deformation of the spacetime, modifying the location of characteristic radii such as the unstable and stable circular orbits, the ISCO, and the separatrix between capture and scattering. To distinguish coordinate effects from
physical changes of scale, we have systematically compared the
coordinate radius $r$ with the invariant areal radius
$R(r)=\sqrt{r^2+A^2}$. Although the coordinate location of the
horizon moves inward with increasing $A$, its areal radius increases.
Moreover, the outward shifts of the unstable circular orbit, the ISCO,
and the scattering threshold persist when expressed in terms of
$R$, while the stable circular orbit considered at fixed angular
momentum moves inward in both descriptions. These results demonstrate
that the displacement of the characteristic orbital scales represents
a genuine geometric effect rather than a coordinate-dependent effect.

Near the critical value $A/m=3\pi/2$, the distinction between
coordinate and invariant radial scales becomes especially important:
while the horizon approaches the coordinate position $r_+=0$, its
areal radius remains finite. The $A\simeq4.71$ configurations shown
in our numerical analysis therefore represent the limiting
black-hole geometry. Beyond this value,  the throat moves outside the horizon and lies in the static (R) region. The spacetime thus develops a wormhole-like structure, and the interpretation of plunge and capture trajectories must be reconsidered accordingly.

In the bounded regime, the perihelion precession was
computed through a perturbative solution of the generalized
Binet equation. The resulting shift contains the standard general
relativistic contribution plus an additional correction proportional
to $A^2/\ell^2$. By confronting the theoretical prediction with the
Solar System data, we obtained the strongest upper bound on the
scalar charge from Mercury, $A\leq1.048\times10^{6}\,{\rm m}$,
indicating that deviations from Schwarzschild geometry must remain
small at planetary scales and that deviations from General Relativity
remain strongly suppressed in the weak-field regime. These results are
consistent with our previous analysis based on null geodesics and
observational signatures \cite{Gonzalez:2025yjm}. To compare this
dimensional bound with the critical ratio associated with the stability
condition discussed in Ref.~\cite{Bronnikov:2012ch}, we use the same
solar mass scale adopted in the perihelion analysis,
$m=m_{\odot}=GM_{\odot}^{\rm phys}/c^2\simeq1476.1\,{\rm m}$.
For a solar-mass source, the critical ratio $A/m=3\pi/2$ therefore
corresponds to $A_{\rm crit}\simeq6.96\times10^{3}\,{\rm m}$.
This value lies well below the strongest Solar-System upper bound,
$A_{\rm max}=1.048\times10^{6}\,{\rm m}$, and is therefore not excluded
by the weak-field constraint derived from perihelion precession.
We emphasize, however, that the Solar-System bound does not by itself
establish the stability of the solution; it only shows that the
particular critical configuration associated with
$A/m=3\pi/2$ remains compatible with the observational bound obtained
here.

In the unbounded regime, the scalar hair modifies the height and position of the effective potential barrier, altering the transition between capture and scattering. 
In particular, the critical angular momentum $L_S$ and the corresponding unstable circular radius increase with $A$, shifting the onset of the scattering region to larger geometric distances. 
Nevertheless, the qualitative structure of timelike geodesics remains continuously connected to the Schwarzschild case in the limit $A \to 0$.

Overall, our results show that regular black holes with phantom scalar hair preserve the global qualitative structure of massive geodesic motion while introducing controlled quantitative deviations encoded in the scalar charge. 
These deviations manifest in orbital stability properties, separatrix structure, and classical tests such as perihelion precession. 
The analysis therefore provides a complementary phenomenological assessment of regular black hole geometries beyond photon observables, extending previous studies of null geodesics to the massive sector.

\begin{acknowledgments}

We thank the anonymous referee for valuable comments and suggestions. Y. V. acknowledges support by the Direcci\'on de Investigaci\'on y Desarrollo de la Universidad de La Serena, Grant No. PR25538511.

\end{acknowledgments}

\end{document}